\documentclass{aa} 

\usepackage{graphicx}
\usepackage{txfonts}
\usepackage[colorlinks=true,citecolor=blue]{hyperref}
\begin{document}

   \title{Nonradial and nonpolytropic astrophysical outflows.}

   \subtitle{X. Relativistic MHD rotating spine jets in Kerr metric}

   \author{
 L. Chantry\inst{1}
    \and
          V. Cayatte\inst{1}
    \and      C. Sauty  \inst{1}
    \and      N. Vlahakis\inst{2}
    \and      K. Tsinganos\inst{2}
          }

   \institute{
 Laboratoire Univers et Th\'eories, Observatoire de Paris - PSL, UMR 8102 du CNRS, Universit\'e Paris Diderot, F-92190 Meudon, 
 France
         \and
Section of Astrophysics, Astronomy and Mechanics, Department of Physics, University of Athens, Panepistimiopolis Zografos, Athens 15783, Greece
\\
        \email{christophe.sauty@obspm.fr}
             }

 \date{Received ....; accepted ....}

 \abstract
{High resolution radio imaging of Active Galactic Nuclei (AGN) have revealed that some sources present motion of superluminal knots 
and transverse stratification of their jet.
Recent observational projects, e.g., ALMA and $\gamma$ - ray telescopes such as HESS and HESS2 (also in the future the CTA) have provided
new observational constraints on the central region of rotating black holes in AGN, suggesting that there is an inner- 
or spine-jet surrounded by a disk wind. This relativistic spine-jet is likely to be composed of electron - positron pairs extracting energy from the black 
hole.}
 {In this article we present an extension and generalization to relativistic jets in Kerr metric of the Newtonian meridional self-similar 
 mechanism of  \cite{ST94}. We aim at modeling the inner spine-jet of AGN as the relativistic light outflow emerging from a spherical 
 corona surrounding a Kerr black hole and its inner accretion disk. 
}
  {The model is built by expanding the metric and the forces with colatitude to first order in the magnetic flux function. As a result of the 
  expansion, all colatitudinal variations of the physical quantities are quantified by a unique parameter. Conversely to previous models, 
  effects of the light cylinder are not neglected.}
 { Solutions with high Lorentz factor are obtained and provide spine-jet models up to the polar axis. As in previous publications, 
 we calculate the magnetic collimation efficiency parameter, which measures the variation of the available energy across the field lines. 
 This collimation efficiency  is an integral of the model, generalizing to Kerr metric the classical magnetic rotator efficiency criterion. 
 We study the variation of the magnetic efficiency and acceleration with the spin of the black hole and show their high sensitivity to this integral.}
 {These new solutions model collimated or radial, relativistic or ultra-relativistic outflows in AGN or Gamma-Ray Bursts (GRB). 
 In particular, we discuss the relevance of our solutions to model the M87 spine-jet. 
 { We study the efficiency of the central black hole spin to collimate a spine-jet and show that the jet power is 
 of the same order with that determined by numerical simulations.}
 %By extending this model to inflow of electron - positron pairs, 
 }


   \maketitle
%
%________________________________________________________________

\section{Introduction}

AGN jets are now recognized to be multi-component outflows related to accretion onto a supermassive black hole (SMBH). 
The outer kiloparsec/megaparsec  scale lobes are fed by a powerful hadronic plasma, which most likely originates from the Keplerian accretion disk 
via the magnetocentrifugal launching mechanism, \cite{BlandfordPayne82}. This hadronic population could be responsible for the 
second component peaking in the $\gamma$-ray band of the spectral energy distribution (SED) of some blazars. As explained in  
\cite{Boettcheretal13}, this is possible if a relativistic jet of protons contributes significantly to the radiative output through 
proton synchrotron emission or photo-pion production. However the purely hadronic synchrotron models for blazars present a 
major problem due to the requirement to have very high powers in the jets. Indeed, hadronic processes are very inefficient (\citealt{Sikora11}). They are energetically 
favorable only for high-frequency synchrotron peak (HSP) blazars, where the jet is likely to be highly magnetized (\citealt{Petropoulou16}). 

The alternative approach to explain the origin of the high energy emission is to consider leptonic models. There, the radiative 
output throughout the electromagnetic spectrum  is assumed to be dominated by leptons both at low and high frequencies. Inverse 
Compton scattering of soft photons (IC) by relativistic nonthermal electrons in the jet is supposed to be the most probable mechanism for 
$\gamma$-ray production,  at least for strong-line and low-frequency synchrotron peak -LSP- blazars. Until recently, leptonic models in 
quasi steady-state were very successful in modeling the SED for almost all classes of blazars (e.g. \citealt{CelottiGhisellini08}). 
However leptonic models with a one-zone component where only one portion of the jet dominates the emission are now 
questioned. Mostly, they fail to produce the extremely high bulk Lorentz factors in a very compact emission region required for ultra-fast 
variability of some TeV blazars (\citealt{Begelman08}). Despite several proposed mechanisms mostly based on small-scale 
inhomogeneities in the jet (see references in \citealt{Vovk15}), it is not yet possible to know the location of the variable $\gamma$-ray 
emission. It may be at the base of the jet or up to parsec-scale distances from the central black hole. Moreover most of the sources in  
which ultra-rapid variability events have been observed are BL Lac objects emitting very high energy radiation.  

For BL Lacs the application of the one-zone synchrotron-self Compton model to the SED implies that the jet is weakly magnetized and 
the emitting region is far from equipartition. Both conditions are required if magnetic reconnection powered emission is at the origin of 
the ultra-rapid variability (\citealt{Tavecchio16}). A most important constraint comes from the launching and the acceleration processes, 
which should lead to nearly equipartition between the magnetic and the kinetic energy fluxes as it can be seen in other blazars. 
In order to unify the BL Lac and radiogalaxy populations, \cite{Ghisellini05} already proposed a structured jet model, with two 
components, a faster core (the spine) surrounded by a slower sheath or layer. This two component model is able to explain 
peculiar features of TeV emitting BL Lacs, such as the absence of fast superluminal components and the presence of a limb-brightened 
radio structure (\citealt{Giroletti04}, \citealt{Nagai14}). \cite{TavecchioGhis16} showed that this structured jet model gives 
an extra source of soft photons intervening in the IC emission for BL Lacs. It is due to the radiative interplay between the two 
components and allows to reproduce the emission in equipartition conditions. This spine-sheath jet structure has also been explored 
in \cite{Sikora16} for strong-line blazars in order to explain why $\gamma$-ray variations are often observed to have much larger 
amplitudes than the corresponding optical variations as well as other observed features of $\gamma$-ray flares.
For the nearby radiogalaxy M87, a direct detection of different motions was possible and an extended mildly relativistic flow is seen 
surrounding a relativistic central jet (\citealt{Mertens16}).

The possible seed photon population at the origin of IC scattering in AGNs is subject to questioning. The dominant seed source 
depends critically on the location of the primary emitting region in the jet (\citealt{Finke16}). Radio Array observations of individual 
blazars are used to localize the $\gamma$-ray emission site among the superluminal components seen along the jet. However, they 
cannot put strong constraints on localizing the emission regions of the high-energy flares. More theoretical and numerical modelings 
and carefully analyzed polarization data are needed (see \citealt{Kravchenko16} and references therein). Note also the treatment 
used in \cite{Hada11} for locating the central black hole relatively close to the jet base-radio core in M87 using the dependence of the core 
position on frequency. There are cases where the spine-sheath model should work. The leptons in superluminal components after
leaving the black hole corona interact with the sheath where IC scattering of photons occurs and produce the observed $\gamma$-ray 
activity (\citealt{Marscher10}; \citealt{Casadio15}).

For AGN jets, two-component models have been theoretically studied in \cite{Soletal89} and further developed by several authors 
(e.g. \citealt{FabianRees95}). In those studies the spine jet is the inner leptonic component (electron-positron pairs), which is self 
collimated inside a hadronic disk wind, and should be separated from the wind by a force-free, empty region. On the other hand, it 
has been shown numerically that the high collimation efficiency of the outer disk wind is sufficient to prevent decollimation from the 
inertia of the inner relativistic plasma, in the case where the initial magnetic field lines going out from the launching region, i.e. 
the black hole corona, are radial (e.g. \citealt{Gracia09}). Moreover synthetic synchrotron maps build from such MHD models can 
reproduce the opening angle of M87 up to a large distance from the black hole, and the change between centre-brightening near the 
core to limb-brightening further away.

The standard accretion-jet model includes an advection-dominated accretion flow (ADAF) which takes place at least in the inner part 
of the accretion disk in AGNs, except for very high accretion efficiency (\citealt{Nar94}).  The presence of this ADAF structure in the disk
center does not prevent the launching of the internal jet as it can be seen from
observations near the black hole for radiogalaxies (see the Faraday rotation measures from observations of M87 in 
\citealt{Feng16}). In the high accretion regime for AGNs, where a standard thin disk is found and powerful outflows are observed, 
both an internal jet and an outer disk wind coexist. 

Numerical simulations in the framework of general relativity have been performed to model the inner jet formation 
assuming a large-scale magnetic field right from the beginning, mostly using monopolar configuration anchored in the ergosphere.   
\cite{Komissarov07} was able to ensure the viability of the Blandford-Znajek mechanism (\citealt{BlandfordZnajek77}) and the 
possibility of building a relativistic particle outflow. Similar relativistic  simulations of jets were obtained by \cite{McKinneyBlandford09} 
where the central Poynting-flux dominated force free jet is self-consistently feeding the funnel of the accretion disk wind created along 
the axis by the very high toroidal field in a magnetically arrested disk (MAD). \cite{McKinney2012} and \cite{Tchek2011} have also derived 
a scaling law between the magnetic flux of the field threading the black hole and the mass accretion rate on the horizon. This has been confirmed 
observationnaly by \cite{Zama2014}. In addition the jet power, directly related to the magnetic flux on the horizon, was calculated in their simulations.
allowing to derive a net flow efficiency between the accretion and the jet. 

Beyond the launching region, the internal structure of 
magnetized and relativistic jets can be probed by RMHD simulations as it has been done by \cite{Marti16}. They characterized the 
internal jet structure of overpressured, steady jets in connection with the dominant energy (internal, rest-mass and magnetic). They 
showed that transverse equilibrium with a significant toroidal magnetic component implies a structure with a central spine and a 
surrounding layer with lower thermal and total pressures. Thus, we have strong clues, both from observations and simulations, that 
relativistic jets have a transverse structure. On the other hand, the jet launching 
mechanism associated with magnetic flux threading the black hole seems to produce only very light spine jets with rarefied gas such 
that the inner outflow is force-free. The inner beam would be so light that it would be invisible at large distances. An alternative issue,  
in order to model the spine jet, is to extend to general relativity the self-similar solutions for jet outflows in young stellar objects.
This model can be derived analytically and used as initial conditions in numerical simulations.

A self-similar model of the spine jet has been proposed by \cite{Melianietal06} for non rotating black holes in Schwarzschild metric, 
extending the previous Newtonian self-similar model of \cite{ST94}. The \cite{Melianietal10} model aimed to explain the 
dichotomy between FRI and FRII sources in terms of magnetic collimation efficiency. Then, \cite{Globusetal14}, have partially 
generalized this model to a Kerr metric. However they fail at giving an exact generalization of the magnetic collimation criterion linked to the efficiency of the magnetic rotator.   
In both cases, the authors assume that the light cylinder is sufficiently far from the spine jet to have its effects neglected. Physically, 
this means that both models cannot produce solutions consistent across the light cylinder. 
  
In the present paper,  we present a new extension of the non relativistic meridional self-similar solutions of \cite{ST94}, both in a 
Schwarzschild and a Kerr metrics.  This model can produce solutions for relativistic jets emerging from a spherical corona surrounding 
the central part of a Kerr black hole and its inner accretion disk.

Conversely to previous models in a Schwarzschild or a Kerr metrics, this new model for a non rotating, as well as for a rotating black hole 
includes a self consistent crossing of the light cylinder in the solution. The Alfv\'en Mach surfaces are not spherically 
symmetric. However, the  Alfv\'en transition surface is spherical and includes the light cylinder. 

In Sec. 2 we summarize the ideal MHD equations in the 3+1 formalism and in Sec. 3 construct the model and derive its 
equations. In Sec. 4 we describe how to solve the equations and illustrate this procedure with various new solutions in a Schwarzschild metric. Then,  
four different solutions in a Kerr metric are presented in Sec. 5 with high Lorentz factor and different geometries potentially applicable to 
AGNs and GRBs.  Some characteristics of those solutions of the model are discussed as well as how the magnetic collimation evolves with the 
rotation of the black hole. 
 
%__________________________________________________________________

\section{Steady axisymmetric relativistic MHD outflows \label{Section2}}
%__________________________________________________________________

 \subsection{Kerr metric} 
The first step in building self similar solutions in relativistic flows is to define the metric. In fact the central massive black hole 
dominates the gravitational field in the near regions and determines completely the metric field. 
Thus, in Kerr metric, the geodesics are defined by,
\begin{eqnarray}
&ds^2 = & -\left(1-\dfrac{r_s r}{\rho^2} \right) c^2{dt}^2 - \dfrac{2r_s r c a}
{\rho^2}\sin^2 \theta \, dt\, d\phi 
\nonumber\\
&&+\dfrac{\rho^2}{\Delta}{dr}^2 +\rho^2{d\theta}^2+\dfrac{\Sigma^2}{\rho^2}\sin^2\theta \,  {d\phi}^2 
\,.
\end{eqnarray}
We have the usual notations of the elements,
\begin{eqnarray} 
&\Delta&=r^2+a^2-r_s r \,, 
\\
&\rho^2&=r^2+a^2\cos^2\theta \,,
\\
&\Sigma^2&=(r^2+a^2)^2-a^2\Delta \sin^2\theta 
\,,
\end{eqnarray}
where $ a=\dfrac{\mathcal J}{ \mathcal{M}c}$ and $r_s=\dfrac{ 2\mathcal{G}\mathcal{M}}{c^2}$.

Note that $\mathcal J$ is the angular momentum of the massive central object, $\mathcal M$ is its mass, $h$ is the lapse function, $
\omega$ is the angular velocity of zero angular momentum observers (ZAMO) and 
we use $a$ for the length-scale related to the angular momentum of the black hole (Kerr scale). 
We can define the dimensionless spin of the black hole $a_H$ in units of the gravitational radius $r_s/2$ such that,
$a_H = 2 a/r_s$. Furthermore,  $\vec{\beta}$ is the shift vector.

The lapse function $h$, the angular velocity $\omega$ of zero angular 
momentum observers (ZAMO) and the shift vector coordinates can be written as:
\begin{eqnarray}
h=\left(1-\dfrac{r_s r}{\rho^2}+\beta^{\phi}\beta_{\phi}\right)^{1/2}=\dfrac{\rho}{\Sigma}\sqrt{\Delta} \,,   \;\;\;\;\;\;& 
\\
\omega=\dfrac{a c {r_s} r}{\Sigma^2}\,,
\,\,\,\,\,\, \beta_{\phi}=-\dfrac{\omega}{c}\varpi^2\,,
\,\,\,\,\,\, \beta^{\phi}=-\dfrac{\omega}{c}\,
\,.  &
\end{eqnarray}
with $\varpi=\dfrac{\Sigma}{\rho}\sin \theta$. The corresponding line elements for the Kerr metric are given in Appendix \ref{appendixA}.

\subsection{Maxwell's equations}

The next step is to define the electromagnetic field in this metric. Using covariant derivatives, we can write Maxwell's equations in Kerr 
space, assuming stationarity and axisymmetry,
\begin{eqnarray}
\nabla\cdot \vec{E} & = & 4\pi \rho_e\,,\label{MGauss}\\
\nabla\cdot \vec{B} & = & 0\,,\label{MFlux}\\
\nabla\times(h\vec{E}) & = & \left( \vec{B}\cdot\nabla\dfrac{\omega}{c} \right) \varpi\vec{\epsilon}_{\phi}\,,\label{MFaraday}\\
\nabla\times(h\vec{B}) & = & \dfrac{4\pi h}{c}\vec{J}-\left(\vec{E}\cdot\nabla \dfrac{\omega}{c}\right)\varpi\vec{\epsilon}_{\phi}
\label{MAmpere}\,,
\end{eqnarray}
where $(\vec{\epsilon}_i)_{i=1...3}$ is the space orthonormal basis. Note that all quantities in the above equations are given in the 
ZAMO frame. 
We can split all vector fields in a poloidal component in the meridional plane and a toroidal one along the
azimuthal direction. The poloidal magnetic field $\vec{B}_{\rm p}$ can be expressed in terms of the magnetic flux function $A$,
\begin{equation}
\vec{B}_{\rm p}=\nabla\times\left(\dfrac{A}{\varpi}\vec{\epsilon}_{\phi}\right)
\label{BPdef}
\,,
\end{equation}
and using Faraday's law,  we get the electric field,
\begin{equation}
\nabla\times\left( h\vec{E}-\dfrac{\omega}{c}\vec{\nabla}A \right)=0\,.
\end{equation}
We can introduce the electric potential,  $\Phi$, but the electric field $\vec{E}$ is not directly proportional to the gradient of the electric 
potential, 
\begin{equation}
h\vec{E}=\dfrac{\omega}{c}\vec{\nabla}A-\vec{\nabla}\Phi \,.
\label{Eexp1}
\end{equation}
The condition of ideal MHD for infinite electrical conductivity leads to, 
\begin{equation}
\vec{E}+\dfrac{\vec{V}\times \vec{B}}{c}=0\,.
\label{icond}
\end{equation}
Note that $\nabla$ is the covariant derivative on a space hyper-surface, see Appendix \ref{appendixB} for the expression of its 
coordinates.

\subsection{Equations of motion}

In the Kerr metric, the 3+1 formalism gives the equation for mass conservation, the Euler equation and the energy conservation, respectively,
\begin{eqnarray}
\label{3+1}
\nabla \cdot (\rho_0 \gamma h \vec{V}) = 0\,,	
\\
\rho_0 \gamma \left(\vec{V}\cdot \nabla \right)\left(\gamma \xi \vec{V}\right)+\rho_0\xi\gamma^2\left[c^2 {\nabla \ln h}
+\dfrac{\varpi\omega V^{\hat{\phi}}}{h}{\nabla \ln \omega }\right]
\nonumber\\
+\nabla P = \rho_e \vec{E}+\dfrac{\vec{J}\times\vec{B}}{c}\,,
\label{EUL}
\\
\gamma^2\rho_0\xi c \left[\vec{V}\cdot {\nabla \ln(\gamma\xi h)}+\dfrac{\omega\varpi V^{\hat{\phi}}}{hc^2} \vec{V}\cdot
{\nabla \ln \omega}\right]  
=  \dfrac{\vec{J}\cdot\vec{E}}{c}\,.
\end{eqnarray}   
Here  $V^{\hat{\phi}}$ is the toroidal component of the bulk flow speed as seen by the ZAMO. The factor $\gamma$ is the bulk Lorentz factor, $\rho_0$ is the mass density and $\xi c^2$ the specific enthalpy measured in the comoving frame of the outflow, that contains kinetic enthalpy of perfect relativistic gas $\xi_K$ and some heating term $Q/c^2$. 

\begin{equation}
\label{TETENT}
\xi=\xi_K+\frac{Q}{c^2}
\end{equation}

For the kinetic enthalpy we use the Taub-Matthews approximation of ideal fluid equation of state, for more details see \cite{1948PhRv...74..328T}, \cite{Melianietal04} and \cite{2005ApJS..160..199M}.

\begin{equation}
\label{ENTKIN}
\xi_K=\frac{5}{2}\left(\frac{P}{\rho_0 c^2}\right)+\sqrt{1+\left(\frac{3P}{2\rho_0 c^2}\right)^2}
\end{equation}

The energy conservation has been derived in the frame of the ZAMO and equivalently the first law of thermodynamics can be obtained 
by projecting the conservation of the energy-momentum tensor 
along the fluid 4-velocity but in the comoving frame. Assuming infinite conductivity, the contribution of the
electromagnetic field is null and only the thermal energy affects the variation of the enthalpy of the fluid, giving,
\begin{equation}\label{FiPr}
{\rho_0 } \left(\vec{V}_{\rm p}\cdot \nabla\right) (\xi c^2)=  \left(\vec{V}_{\rm p}\cdot \nabla\right)P
\,.
\end{equation}

\subsection{Constants of motion}

Under the assumptions of steadiness and axisymmetry, the magnetohydrodynamic equations in 
general relativity can be partially integrated
to yield several field/streamline constants (\citealt{beskin10}). We already deduced those constants 
including the magnetic field with the same formalism in \cite{cayatteetal14}. Here we present the derivation of
the equations, following the notations of \citealt{Tsing82}, in order to compare with previous self-similar models, e.g.,  
\cite{Melianietal06} and give the choice of the first integrals.

Steady and axisymmetric flows are characterized by a function $A$ that defines the geometry 
of the magnetic flux surfaces. In the poloidal plane, field lines are lines of constant magnetic flux $A$ and first integrals
will be functions of $A$, among which the mass flux $\Psi$. The poloidal velocity can be expressed in terms of $\Psi$,
\begin{equation}\label{VitPol}
4 \pi \rho_0 \gamma h \vec{V}_{\rm p}=\nabla\times\left(\dfrac{\Psi}{\varpi}\vec{\epsilon}_{\phi}\right)
\,.
\end{equation}
The frozen-in condition for ideal MHD flows, gives in the toroidal  
direction, combined  with Eq. (\ref{BPdef}),
\begin{eqnarray*}
4 \pi \rho_0 \gamma h \vec{V}_{\rm p}=\Psi_A \vec{B}_{\rm p}
\,,
\end{eqnarray*}
where $\Psi_A\equiv{\displaystyle{\rm d}\Psi}/{\displaystyle{\rm d}A}$ is the magnetic to mass flux ratio. 

The poloidal components of the law of flux freezing (Eq. \ref{icond}) give in turn the iso-rotation law,
\begin{equation}
\Omega - \omega = \dfrac{hV^{\hat{\phi}}}{\varpi}-\dfrac{\Psi_A B^{\hat{\phi}}}{4\pi\rho_0\gamma \varpi}
\label{Omega}
\end{equation}
where $\Omega(A)\equiv c{\displaystyle{\rm d}\Phi}/{\displaystyle{\rm d}A}$ is the isorotation frequency, which is constant along each 
magnetic flux tube. 

By integrating the Euler equation in the toroidal direction, we get the conservation of the angular momentum flux $L(A)$, \begin{equation}
 L= \varpi \left( \gamma \xi V^{\hat{\phi}}-\dfrac{h B^{\hat{\phi}}}{\Psi_A}\right)
\label{L}\,.
\end{equation}

The last equation to integrate is the energy conservation. In other words, we may take the Euler equation projected along the time axis 
of the 3+1 decomposition, and integrate it under the hypothesis of steadiness,
\begin{equation}
\mathcal{E}-L\omega=\gamma\xi h c^2- \dfrac{h\varpi (\Omega-\omega)}{\Psi_A} B^{\hat{\phi}}
\label{E}\,.
\end{equation}

\subsection{Toroidal fields}
\label{Toroidalfields}
Using the three last integrals of motion, we may express the toroidal components of the velocity and the magnetic fields and the 
enthalpy density  as functions of these first integrals and the poloidal components. Using the standard procedure of inversion we get, 

\begin{eqnarray}
\varpi \dfrac{h B^{\hat{\phi}}}{\Psi_A} = \dfrac{L\left[ h^2 c^2
+ \varpi^2 \omega (\Omega-\omega) \right] - \mathcal{E}\varpi^2 (\Omega-\omega)
         }
{ \left( M_{\rm Alf}^2-h^2 \right)c^2 + \varpi^2 (\Omega-\omega)^2 }
 \label{INV1}\,,
 \\
\varpi  \gamma \xi V^{\hat{\phi}} = 
\dfrac{M_{\rm Alf}^2Lc^2-(\mathcal{E}-L\Omega)\varpi^2(\Omega-\omega)}
{ \left( M_{\rm Alf}^2-h^2 \right)c^2  + \varpi^2 (\Omega-\omega)^2}
 \label{INV2} \,,
 \\
\gamma h \xi =  
\dfrac{M_{\rm Alf}^2(\mathcal{E}-L\omega)-h^2(\mathcal{E}-L\Omega)}
{\left( M_{\rm Alf}^2-h^2 \right)c^2  + \varpi^2 (\Omega-\omega)^2}
\,,
 \label{INV3}
\end{eqnarray}    
where we have defined the poloidal Alfv\'en Mach number, 
\begin{equation}\label{defMa}
  M_{\rm Alf}^2=h^2\dfrac{{V_{\rm p}}^2}{V_{\rm Alf}^2}=\dfrac{4{\pi}h^2\rho_0\xi\gamma^2{V_{\rm p}}^2}{{B_{\rm p}}^2}=\dfrac{\xi 
  {\Psi_A}^2}{4\pi\rho_0}
 \,.
\end{equation}
This definition of the poloidal Alfv\'en Mach number is 
consistent with the definition used by \cite{Melianietal06} and includes the lapse function. This is also the definition taken by 
\cite{Breitmoser2000} because the velocity, $hV_{\rm p}$, calculated with the universal time is continuous across the event horizon. 

The numerator and denominator of Eq.  (\ref{INV1}) are zero at the Alfv\'en transition surface, if the following two equations are satisfied,
\begin{equation}
\label{DENalf} 
\left. M_{\rm Alf}^2 \right|_{\rm a}=h_{\rm a}^2\left[1-\frac{\varpi_{\rm a}^2(\Omega-\omega_{\rm a})^2}{h_{\rm a}^2 c^2}\right]\,, \\
\end{equation}
\begin{equation}
\frac{\varpi_{\rm a}^2(\Omega-\omega_{\rm a})^2}{h_{\rm a}^2 c^2}=
\dfrac{L (\Omega-\omega_{\rm a})}{(\mathcal{E}-L\omega_{\rm a})}
\,.
\label{NUMalf}
\end{equation}
The denominators of Eqs.  (\ref{INV2})  and (\ref{INV3}) are identical to the one of Eq.  (\ref{INV1}) and there numerators are a linear
combination of Eq. (\ref{DENalf}) and Eq.  (\ref{NUMalf}). So the numerators and the denominators of Eqs.  (\ref{INV2})  and  
(\ref{INV3}) are also zero at the Alfv\'en transition surface.

We can reformulate the above equations by changing variables. We rescale the cylindrical radius with $ch/(\Omega-\omega)$
leading to the dimensionless cylindrical radius $x$ and introduce the parameter $x_{\rm MR}$,
\begin{eqnarray}
x=\dfrac{\varpi (\Omega-\omega)}{h c}\,,
&x_{\rm MR}^2=\dfrac{L (\Omega-\omega)}{(\mathcal{E}-L\omega)} \,.
\end{eqnarray}

Hence we can write,
\begin{eqnarray}
 B^{\hat{\phi}} = \dfrac{- (\mathcal{E}-L\omega)\Psi_A}{c x}
\dfrac{x^2 -  x_{\rm MR}^2}
{ M_{\rm Alf}^2-h^2 \left(1-x^2 \right)},
 \label{INV4}
 \\
h\gamma \xi \frac{V^{\hat{\phi}}}{c} = \dfrac{(\mathcal{E}-L\omega)}{c^2 x}
\dfrac{M_{\rm Alf}^2 x_{\rm MR}^2 - (1-x_{\rm MR}^2) h^2 x^2}
{ M_{\rm Alf}^2-h^2 \left(1-x^2 \right)},
 \label{INV5}
 \\
\gamma h \xi =  \dfrac{(\mathcal{E}-L\omega)}{c^2}
\dfrac{M_{\rm Alf}^2-h^2(1-x_{\rm MR}^2)}
{ M_{\rm Alf}^2-h^2 \left(1-x^2 \right)}
 \label{INV6}
\,,
\end{eqnarray} 

The second condition, Eq. (\ref{NUMalf}), at the Alfv\'en transition surface becomes, keeping the first one unchanged,
\begin{eqnarray}
\left. x^2 \right|_{\rm a} =\left. x_{\rm MR}^2\right|_{\rm a}
\,.
\label{NUMalf2}
\end{eqnarray}

In Kerr metric the parameter $x_{\rm MR}^2$ is an extension of $x_{\rm A}^2$ defined by \cite{Melianietal06}. It measures the amount 
of energy carried by the electromagnetic field. This is  the energy flux of the magnetic 
rotator ({\bf MR}) divided by the total energy flux of the outflow in the co-rotating frame, 
$\mathcal{E}-L\omega$. This new parameter $x_{\rm MR}$, conversely to the previous $x_{\rm A}$,  is not any more constant along a field line, 
since $\omega$ is not an integral of the motion.

%__________________________________________________________________

\section{Model equations}

%__________________________________________________________________

\subsection{Angular expansion}

The MHD equations and the metric constitute a coupled set of highly non linear equations that cannot be solved analytically. 
The approach followed so far for Newtonian flows has been to look for solutions with separable variables in the frame of  
self-similarity. However, this technique cannot be applied in the frame of general relativity, due to the complexity of the metric even for the simpler  
cases of a Schwarzschild, or a Kerr metric. Instead, we may model the jet close to its symmetry axis, i.e.,  to describe the spine jet, by expanding all 
variables with $\sin \theta$ to second order. 
 
 Along the polar axis where $\varpi$ and $\theta$ go to zero, we may define the spherical Alfv\'en radius, to be the distance  $r_\star$ from the center 
 where the Alfv\'en transition surface condition, $M_{\rm Alf, \theta=0}^2=h_\star^2$, applies. The subscript $\star$ denotes the 
value of a physical quantity at the Alfv\'en transition surface, along the polar axis. We shall use this location to write all our 
quantities in dimensionless form.  Thus, the  dimensionless spherical radius is,
\begin{equation}
R=\frac{r}{r_\star}
\,.
\end{equation}

At $R=1$, the velocity is $V_\star$, the magnetic field $B_\star$, the density $\rho_\star$, the enthalpy $\xi_\star$  and the lapse 
function $h_\star$. Because of the Alfv\'en transition along the polar axis, we have,
\begin{equation}
B_\star^2=4\pi{\gamma_\star}^2\rho_\star \xi_\star {V_\star}^2
\,.
\end{equation}  

Thus the dimensionless magnetic flux function $\alpha$ is defined as,
\begin{equation}
\alpha=\frac{2}{{r_\star}^2 B_\star} A
\end{equation}

Moreover we can expand to the second order the metric of the system in dimensionless form
using the characteristic dimensions of the system defined in the previous section. This introduces the two following new parameters,
\begin{equation}
\mu=\frac{r_s}{r_\star}
\,,
\qquad  \qquad
l=\frac{a}{r_\star}  = \frac{\cal J}{{\cal M} c r_\star}\Rightarrow \frac{2a}{r_\star} =\frac{2l}{\mu}
\,,
\end{equation}
which are respectively the Schwarzschild radius in units of the Alfv\'en radius and the dimensionless black hole spin. %\\ DEFINITION OF ${a}$ ?} 

Another dimensionless parameter is needed to describe the gravitational potential, as in the classical model. 
This parameter $\nu$ represents the escape speed at the Alfv\'en point along the polar axis in units of $V_\star$.  
Then, the value of $V_\star$ is fixed by the following condition, 
\begin{equation}
\nu=\frac{V_{esc, \star}}{V_\star}=\sqrt{\frac{2\mathcal{G}\mathcal{M}}{r_\star V_\star^2}}
 \quad  \Rightarrow  \quad
V_\star^2=\frac{\mu}{\nu^2} c^2
\,.
\label{Vstar}
\end{equation}

Thus, to second order in $\sin \theta$ the ZAMO angular velocity and the lapse function are written as,		
 \begin{eqnarray}
\omega & = & \frac{lc\mu R}{r_\star(R^2+l^2)^2} \left(1+\frac{l^2 {h_z}^2}{R^2+l^2}\sin^2\theta \right) \\
h_{\,} & = & \sqrt{1-\frac{\mu R}{R^2+l^2}}\left(1-\frac{\mu l^2 R}{2(R^2+l^2)^2}\sin^2\theta \right)
\,.
\label{OHDEF}
\end{eqnarray}    

In order to simplify our notation, we define the lapse function along the polar axis,
 \begin{equation}
h_z(R) = h(R,\theta=0)= \sqrt{1-\frac{\mu R}{R^2+l^2}}
\label{hzdef}
\end{equation}    	
 and the polar shift of the metric, 
 \begin{equation}
\omega_z(R)=\omega(R,\theta=0)=\frac{lc\mu R}{r_\star(R^2+l^2)^2}
\,.
\label{omzdef}
\end{equation}  	  
See Appendix \ref{appendixA} for details.

It will be useful to introduce the dimensionless polar shift function (see also Eq. \ref{Omom}),
\begin{equation}
\overline{\omega}_z(R)=\frac{\omega_z r_\star}{V_\star h_\star}=\frac{l \sqrt{\mu}\nu R}{h_\star(R^2+l^2)^2}
\,.
\label{ombzdef}
\end{equation} 

We also expand the magnetic flux function to second order in $\sin\theta$. The magnetic flux is an even function which is zero along the polar axis because of 
axisymmetry and of the symmetry around the equatorial plane. Thus all odd orders are zero and the first non vanishing even order is 
the second order in colatitude. If we keep the lowest order in the expansion we get,
\begin{equation}
\alpha(R,\theta)= f (R) \sin^2\theta 
\,,
\end{equation}
where $f$ is the inverse of the classical expansion factor for solar coronal holes (see \citealt{TS92a}). 
This expansion similarly to the classical self similar model of \cite{ST94}, is equivalent to an hypothesis of separation of the variables 
in the magnetic flux function. 

Thus from Eq. (\ref{OHDEF}), the cylindrical radius can be also seen as an expansion in the magnetic flux. This
 is physically more meaningful as the magnetic flux is constant on a given mass flux tube. Moreover, several free integrals solely 
depend on this magnetic flux.
We define the dimensionless cylindrical radius $G$ in units of the polar Alfv\'en radius as,
\begin{equation}
G(R)=\sqrt{\frac{R^2+l^2}{f(R)}}
\,.
\end{equation}

 The cylindrical radius can be written in the various following forms,
 \begin{eqnarray}
 \varpi^2&=r_\star^2(R^2+l^2)\sin^2\theta 
 &={r_\star}^2 G^2 \alpha =G^2\varpi_a^2
 \,,
\end{eqnarray}

We can also write the metric as an expansion in $\alpha$ (see also Appendix \ref{appendixA}),
 \begin{eqnarray}
\omega & = & \frac{lc\mu R}{r_\star(R^2+l^2)^2} \left(1+\frac{l^2 {h_z}^2 G^2}{(R^2+l^2)^2}\alpha \right)\,, \\
h_{\,} & = & \sqrt{1-\frac{\mu R}{R^2+l^2}}\left(1-\frac{\mu l^2 R G^2}{2(R^2+l^2)^3}\alpha \right)
\,.
\label{OHDEF}
\end{eqnarray}    
%%%%%%%%%%
Of course we can always reverse our point of view and go back to the expansion in $\theta$. This would be the case if we want to use 
the steady analytical solution as initial conditions for numerical simulations.  

We can parametrize the geometry of the flux tubes with the logarithm derivative of $f$ denoted $F$,
\begin{equation}
 F=\frac{{\rm d} \ln f}{{\rm d} \ln R}=2\left(\frac{R^2}{R^2+l^2}-\frac{{\rm d} \ln G}{{\rm d} \ln R}\right)
 \,.
\label{FG2}
\end{equation}

The angle $\chi$ of the magnetic poloidal field line with the radial direction (see \citealt{STT99}) is given in our metric by,
\begin{equation}
\tan \chi=\frac{\sqrt{R^2+l^2-\mu R}}{2 R}  F\tan \theta 
\,.
\end{equation}

\subsection{Choice of the Alfv\'en surface and pressure}

We can expand all physical quantities to the first order in $\alpha$. Thus the Alfv\'en number is given by, 
\begin{equation}
M_{\rm Alf} = M(R)\left(1+M_1(R) \alpha\right)
\label{Alf}
\,.
\end{equation}
Contrarily to previously self-similar models, the Alfv\'en number cannot be spherically symmetric because of the presence 
of the cylindrical radius in units of the "light cylinder" $x$, in the numerator and the denominator 
of Eqs. (\ref{INV4}),(\ref{INV5}),(\ref{INV6}). This is induced by the regularity conditions, Eqs (\ref{DENalf}) and (\ref{NUMalf}), and the 
sphericity of the Alfv\'en surface. The surface $x=1$ is the so called outer "light cylinder". Of course this surface may not be exactly cylindrical if $x$ depends also on $\alpha$, which may be the case for instance close to the black hole where $\omega$ has a strong depence on $\alpha$ or if $\Omega$ is not constant with $\alpha$. So this is rather a light surface but for the sake of simplicity we shall call it "light cylinder" with quote in the rest of the text.

Similarly the pressure can be expanded to first order,
\begin{equation}
P(R,\alpha)=P_0+\frac{{\gamma_\star}^2{\rho_0}_\star\xi_\star{V_\star}^2}{2}\Pi (R)\left(1+K(R) \alpha\right)
\,,
\label{Pdef}
\end{equation}    
where $P_0$ is a constant. \\
In order to simplify and as a first step, we assume for both equations 
that the radial dependence of the non polar component of the Alfv\'en number and the pressure are simply constant, 
$M_1(R)=m_1={\rm cst}$, $K(R)=\kappa={\rm cst}$. Thus,
\begin{equation}
M_{\rm Alf} =M(R)\left(1+m_1 \alpha\right)
\,.
\label{MAlf}
\end{equation}
Note that we have $m_1=0$ in previous models, see \cite{Melianietal06} and \cite{Globusetal14}. 

\subsection{Choice of the free integrals}
\label{Choiceoffreeintegral}

Free integrals are also expanded to the first order in the magnetic flux. 
The mass to magnetic flux ratio is similar to the one in the classical case, 
expanded as,
\begin{equation}
{\Psi_A}^2(\alpha)=\frac{4\pi{\rho_0}_\star h_\star^2}{\xi_\star}(1+\delta \alpha)
\,.
\label{defPsiA}
\end{equation}
where $\delta$ is a free parameter describing the deviations from spherical symmetry of the ratio number density/enthalpy as in 
\cite{Melianietal06} and not of the density itself, conversely to \cite{ST94}.

The total angular momentum loss flux density is given by,
\begin{equation}
\label{AMF1}
\vec{J}=\gamma \rho_0 L h \vec{V}_{\rm p}=\frac{L \Psi_A}{4 \pi}\vec{B}_{\rm p}
\, .
\end{equation}
Thus it is natural to expand the quantity $L\Psi_A$ rather than $L$ itself. $L\Psi_A$ is also the poloidal current density along the polar 
axis and writes as,
\begin{equation}
\label{AMF2}
L\Psi_A=\lambda h_\star B_\star r_\star \alpha
\,.
\end{equation}
The isorotation law can be expanded to first order as well as the total energy,
\begin{equation}
\Omega =  \Omega_\star(1+w_1 \alpha)  \,,
\label{Omom}
\end{equation}
and
\begin{equation}
{\cal E}  = {\cal E}_\star (1+e_1\alpha)
\,,
\label{e_E}
\end{equation}
where we see from Eq. (\ref{E}) that ${\cal E}_\star=h_\star\gamma_\star\xi_\star c^2$.

Although we have some freedom with the choice of $w_1$ and $e_1$, we could choose $e_1=0$ and $w_1=-\delta/2$ to restrict 
ourselves to 
the values of the previous models, in particular in Schwarzschild metric, see \cite{Melianietal06} and \cite{Melianietal10}. 
In fact, the isorotation fonction $\Omega$ does not need to be expanded beyond the zeroth order term 
because $\Omega$ always appears multiplied by another quantity as in $(\Omega-\omega)\varpi$ or $L\Omega$.
Thus, the value of $w_1$ is free and does not affect the solution. 
Conversely the value of $e_1$ affects the whole dynamics and we shall study the effects of its variation in a futur publication.
We already discussed the fact that taking a weak dependence of $\Omega$ on $\alpha$ has the advantage to minimize the variation 
of the "light cylinder" near the base of the jet. Thus for the sake of simplicity, we shall study here the case where 
$e_1=0$ and $w_1=0$.

\subsection{Constraints on the Alfv\'en Mach number, the isorotation law and the angular momentum flux}

The value of $m_1$ is, in fact, determined by the prescription to cross the Alfv\'en transition surface. In order for the denominator in  Eqs. 
\ref{INV1}, \ref{INV2} and \ref{INV3} to vanish at the Alfv\'enic transition, the two following relations given in Eqs. \ref{DENalf} and 
\ref{NUMalf} must be fulfilled. They can be expanded to first order. For the first regularity condition we get,
\begin{eqnarray}
\left. M_{\rm Alf} \right|_{\rm a} =  h_\star(1+m_1 \alpha) \:  \: \:    {\rm with}    \: \: \:
m_1=-\frac{\mu}{2}\left( \frac{\lambda^2}{\nu^2}  + \frac{l^2}{(1+l^2)^3}\right)
\label{m1alfv}
\end{eqnarray}
The first term in the right part of Eq. \ref{m1alfv} is due to the "light cylinder", and the second one to the non sphericity 
of the gravitational field in Kerr metric. 
$m_1$ is negligible whenever the rotational speed $\lambda V_\star$ is sub-relativistic and either the $\mu$ parameter or the angular momentum of the black 
hole are negligible too. 
Since $m_1<0$, there is a limiting field line where we have $M_{\rm Alf}=0$ since the magnetic flux increases going out from the polar 
axis.

To apply the second regularity condition we use the numerator of Eq. \ref{INV1} and we get to the first order in $\alpha$:
\begin{equation}
\Omega_\star-\omega_\star=\frac{\lambda V_\star h_\star}{r_\star}\label{M1O}
\,.
\end{equation}
Thus we can write,
\begin{eqnarray}
\varpi(\Omega-\omega) &=&  {G(R)\sqrt{\alpha}}{\lambda V_\star h_\star \Lambda(R)}\,,
\end{eqnarray}
where,
\begin{eqnarray}
\Lambda(R) &=& \left[1+\frac{\sqrt{\mu} \nu l}{\lambda h_*}\left(\frac{1}{(1+l^2)^2}-\frac{R}{(R^2+l^2)^2}\right)\right]
\end{eqnarray}

The regularity conditions on the Alfv\'en surface fixes the value of $m_1$. Thus the critical Alfv\'en surface is a sphere like in previous 
meridional self-similar models. Beware though that the Alfv\'en transition surface is a {\bf generalized} or  {\bf modified} Alfv\'en surface as it takes 
into account the modification by the "light cylinder" . 

Simultaneously, surfaces of constant Poloidal Alfv\'en Mach Number, $M_{\rm Alf}=$const. , see Eq. (\ref{MAlf}),
are not spherical surfaces conversely to the one defined by \cite{Melianietal06}. Two effects modify it, first the "light cylinder" effect, 
which was neglected in \cite{Melianietal06} and \cite{Globusetal14}, and second the frame dragging effect (Lense-Thirring).

\subsection{Expansion of the velocity and magnetic fields}

The model is obtained using an expansion to the second order for $\sin \theta$ in the Euler equation.  
Due to axisymmetry, first order terms are zero along $r$ and $\phi$ while the antisymmetry along $\theta$ gives that the zeroth and 
second orders are null along the colatitude. 

Then, it gives for the poloidal velocity field,
\begin{eqnarray}
V^{\hat{r}}  &=&  \frac{V_\star M^2}{h_\star^2 G^2} \left\{
1+\sin^2\theta  \left[ \frac{1}{2}\left(\frac{l^2 h_z^2}{R^2+l^2}-1\right)\right. \right.
\nonumber\\
&+& 
 \left. \left.
\frac{R^2+l^2}{G^2}\left(\frac{\lambda^2\mu}{\nu^2}\left(\frac{\Lambda^2 N_B}{D}+\frac{\overline\omega_z }{\lambda}\right)-e_1-\frac{\delta}{2}+2 m_1\right)\right] \right\}
\nonumber\\
V^{\hat{\theta}} &=&  -\frac{V_\star h_z M^2 \sqrt{R^2+l^2} F}{2 h_\star^2 R G^2}\sin\theta
\end{eqnarray}
And for the poloidal magnetic field, we get: 
 \begin{eqnarray}
B^{\hat{r}} & = & \frac{B_\star}{G^2}\left[1+\frac{1}{2}(\frac{l^2 h_z^2}{R^2+l^2}-1)\sin^2\theta\right]
\\
B^{\hat{\theta}} & = & -\frac{B_\star h_z F \sqrt{R^2+l^2}}{2 G^2 R}\sin\theta
\,,
\end{eqnarray}
Now from the  Eqs.(\ref{INV1}) and (\ref{INV2}), we can calculate to the first order in $\sin \theta$ the toroidal components of fields,
\begin{eqnarray}
V^{\hat{\phi}}  =  -\frac{\lambda V_\star h_z \Lambda N_V}{h_\star G^2 D} \sqrt{R^2+l^2} \sin\theta \\
B^{\hat{\phi}}  =  -\frac{\lambda B_\star h_\star \Lambda N_B\sqrt{R^2+l^2}}{h_z D G^2} \sin\theta
\,,
\end{eqnarray}
where the functions $N_V$, $N_B$ and $D$ have been generalized,
\begin{eqnarray}
N_V=\frac{M^2}{h_*^2 \Lambda}-G^2 \\
N_B=\frac{h_z^2}{h_*^2 \Lambda}-G^2 \\
D=\frac{h_z^2-M^2}{h_*^2}
\end{eqnarray}

\subsection{Expansion of the enthalpy, densities and electric field}

We used Eq. (\ref{INV3}) to deduce the enthalpy,
\begin{equation}
\gamma h \xi c^2  =  \gamma_\star h_\star \xi_\star c^2\left[1+\alpha\left(e_1-\frac{\lambda^2 \mu}{\nu^2}
\left(\frac{\Lambda^2 N_B}{D} 
+\frac{\overline\omega_z }{\lambda}\right)\right)\right]
\,,
\end{equation}
and the mass density is given by, 
\begin{eqnarray}
\gamma^2 \rho_0 \xi =  \gamma_\star^2 {\rho_0}_\star \xi_\star\frac{h_\star^4}{h_z^2 M^2}\left[1+\sin \theta^2 
\left\{
\frac{\mu l^2 R}{(R^2+l^2)^2}
\right.
\right.
\nonumber
\\
\left.
\left.
+ \frac{R^2+l^2}{G^2}\left(2e_1-2m_1+\delta-\frac{2\lambda^2\mu}{\nu^2}\left(\frac{\Lambda^2 N_B}{D}+\frac{\overline\omega_z}
{\lambda}\right)
\right)
\right\}
\right]
\,.
\end{eqnarray}

In GRMHD, we also need the expressions of the electric field and the charge density.  The electric field is a second order term for the 
radial component and a first order term for the $\theta$-component,
\begin{eqnarray}
E^{\hat{r}}=-\frac{\lambda V_\star h_\star B_\star}{2 c}\frac{(R^2+l^2)F\Lambda}{ R G^2} \sin^2\theta\\
E^{\hat{\theta}}=-\frac{\lambda V_\star h_\star B_\star}{c}\frac{\Lambda \sqrt{R^2+l^2}}{h_z G^2}\sin\theta
\,.
\end{eqnarray}
Using Maxwell-Gauss Eq. \ref{MGauss}, we calculate  the charge density from the divergence of the above electric field, to zeroth order only,
\begin{equation}
\rho_e=-\frac{\lambda V_\star B_\star h_\star}{2\pi r_\star c}\frac{\Lambda}{h_z G^2}
\,.
\label{dense}
\end{equation}	
With all these quantities we are able to expand the Euler equation. The radial component is expanded to the second order and the 
colatitude component to the first order. From the expansion of poloidal components in the Euler equation (Eq. \ref{EUL}) and using Eq. 
(\ref{FG2}), we can reverse the system to get the equations of the model (see Appendix  \ref{appendixC} for details).   

\subsection{"Light cylinder"}

The rescaling value $ch/(\Omega-\omega)$ for the cylindrical radius used in Eqs. (\ref{INV4} -  \ref{INV6}) has been defined by \cite{Melianietal06} 
as the "light cylinder". It is a surface of revolution $\Sigma_{\rm LC}$  where,
\begin{equation}
x^2=\frac{\varpi^2(\Omega-\omega)^2}{h^2c^2}=1
\,.
\end{equation}
On the "light cylinder", the electric field $\mid \bf E \mid$ is equal to the poloidal component of the magnetic field 
$\mid \bf B_{\rm p} \mid$.
In the present publication, $\Sigma_{\rm LC}$ designed the external "light cylinder", i.e. $x=+1$, though this is not a cylinder as explained earlier, strictly "light cylinder"speaking, but a surface of revolution. 
This external "light cylinder" is outside  the Alfv\'en surface since the denominator 
of Eqs. (\ref{INV4} -  \ref{INV6}), equals to $M_{Alf}^2$  on the "light cylinder", is negative before crossing the Alfv\'en transition surface and 
positive after crossing it.
At large distance in the jet, the lapse function $h$ goes to unity  and  $\Omega-\omega$ tends to $\Omega$ which is assumed constant in our model. Thus, 
$\Sigma_{\rm LC}$ is located on a constant cylindrical radius along the $\it{z}$ axis, becoming a real cylinder. 

From the iso-rotation law, we get,
\begin{equation}\label{RotationSpeed}
\frac{V^{\hat{\phi}}}{c}=x + \frac{\Psi_A B^{\hat{\phi}}}{4\pi \rho_0 \gamma h c} = x + \frac{V_{\rm p}}{c} \frac{B^{\hat{\phi}}}{B_{\rm p}}
\,.
\end{equation}

As in special relativity, the second term of Eq. \ref{RotationSpeed} cannot be neglected in the vicinity of the "light cylinder".  The sign of 
$B^{\hat{\phi}}$ is such that $V^{\hat{\phi}}$ always remains less than the speed of light (\citealt{Vlah15}). Moreover after crossing 
the "light cylinder" one of the two following conditions must be fulfilled. Either, we have
$\mid B^{\hat{\phi}} \mid$ >> $B_{\rm p}$ or $V_{\rm p}$ >> $V^{\hat{\phi}}$, or both.  

The term $x$ was neglected in the equation of the previous relativistic meridional-self-similar models, \cite{Melianietal06, 
Globusetal14}. Hence, these models could not produce jets crossing the "light cylinder". Conversly, in this model this quantity is taken into account. We 
assume an expansion in $\sin(\theta)$ of this quantity. 

Contrary to the two previous models we can choose the dependence of the isorotation frequency 
with the magnetic flux (see discussion on Eq. \ref{Omom}) and this choice will not affect the solution. 
Thus, if $\Omega$ does not depend strongly on the magnetic flux $A$, even at the base of the jet, the ratio $h/(\Omega-\omega)$ will be nearly constant. 
The reason is that $\varpi$ is larger than the Alfv\'en radius which is at least few times the Schwarzschild radius. As a consequence, 
the departure of $\Sigma_{\rm LC}$ from a real cylinder is unnoticeable.

\subsection{Domain of validity}

The equations of the model are the result of an inversion of the expanded conservation equations. So it will be useful to quantify the 
relative error of the expansion we made, in order to analyze properly our results and to get the domain of validity of these results. To 
have an idea of the domain of validity, we will quantify the rest in the expansion of the momentum equation, for each force 
$\mathbf{F}^i(R,\sin \theta)$,
\begin{equation}
\mathbf{F}^i(R,\sin \theta)=\mathbf{F}^i_0(R)+\mathbf{F}^i_1(R)\sin \theta+\mathbf{F}^i_2(R)\sin^2 \theta
+\mathbf{R}^i(R,\sin\theta)\sin^3\theta
\,,
\end{equation}
where $\mathbf{F}$ is one of the following forces, gravitational, centrifugal, inertial, electric force or magnetic pressure etc.... We 
define a new function in order to map the relative error
\begin{equation}
\mathbf{R}^i(R,\sin\theta)\underset{\theta \rightarrow 0}{\sim} \mathbf{g}^i(R)
\,,
\end{equation}

For example, in case of the electric force, we get in Schwarzschild metric, assuming solid rotation ($w_1=0$),
\begin{eqnarray}
\mid \mathbf{R}_{El}(R,\sin\theta) \mid = \frac{ B_\star^2 }{4\pi r_\star}\frac{\lambda^2  h_\star^2 \mu}{\nu^2}
\frac{R}{h_z^2  G^4} \left( \frac{F^2 h_z^2}{ 2}  + \frac{F h_z^2}{2} - 3 + \frac{dF}{dR}\right)
\nonumber \\   \times \sqrt{1 + \sin^2\theta \left( \frac{F^2 h_z^2}{4}-1\right)}
\,,
\end{eqnarray}

The relative error on the electric force which tends to zero in the asymptotic regime of cylindrical jets, is defined as,
\begin{equation}
err=\frac{\mid\mathbf{R}^{El}(R,\sin\theta)\sin^3\theta \mid}{\mid \mathbf{F}^{El}(R,\sin \theta)\mid }
\end{equation}

\begin{figure}[h]	
\centering
\includegraphics[width=9.5cm]{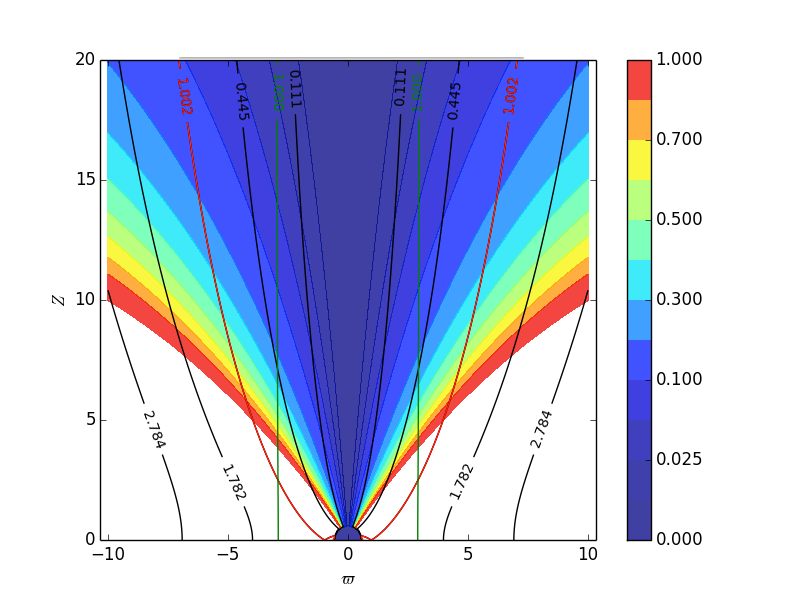}
\caption{Relative error on the electric force for a recollimating oscillating solution in Kerr metric ($\mathbf{K1}$, see Sec.
\ref{RecOscSolK}). 
Color isocontours correspond to the relative error in the electric force.  Field lines anchored into the black hole magnetosphere 
and in the accretion disk are plotted in black solid lines. The limiting field line between the inner jet coming from black hole corona and the outflow 
outgoing from the accretion disk is plotted in red. The "light cylinder" is indicated by a green solid line. 
The cylindrical radius and the distance above the equatorial plane are in units of Schwarzschild radius.}   	
\label{REKMS}
\end{figure}

Even at the base of the jet this error can be reduced as it can be seen in Fig. \ref{REKMS} for the solution in Kerr metric presented in 
Sec. \ref{RecOscSolK} when the co-latitude is less than $30$ degrees.
To get an estimate of the error in the expanded forces, we should add all relative error terms or take the largest one.
This gives an estimate of the domain of validity of the solutions for a given set of parameters. We postpone the full error analysis for a future paper. 

\subsection{The magnetic collimation efficiency, $\epsilon$}

By writing the first law of thermodynamics in the frame of the fluid along streamlines of an axisymmetric flow, 
we can construct a constant of the motion like in the classical case. The first law of thermodynamics reduces to the adiabatic law if  the heating 
is included in some effective enthalpy (see  Eq. \ref{FiPr}). Thus $(\xi c^2)$ is an effective specific 
enthalpy like for polytropic flows where the enthalpy also hides the heating (cf. \citealt{ST94}) but generalized for relativistic outflows (see Eq. \ref{TETENT}).
Using Eq.  (\ref{defMa}), we can rewrite the first law of thermodynamics in the following form,
\begin{equation}
\xi \Psi_A^2 c^2 \left.\frac{d\xi }{dR}\right|_{\alpha=cst}=4 \pi M_{Alf}^2 \left.\frac{d P}{dR}\right|_{\alpha=cst}
\,.
\label{FP1}
\end{equation}
As the magnetic to mass flux ratio and the total energy flux are constant along each streamline, it is equivalent to,
\begin{equation}
\left.\frac{d (\Psi_A^2\xi^2 c^2)}{dR}\right|_{\alpha=cst}=8\pi M_{Alf}^2 \left.\frac{d P}{dR} \right|_{\alpha=cst}
\,.
\label{FP2}
\end{equation}
Note that  $\Psi_A \xi c^2$ is proportional to the thermal energy. If we write 
\begin{equation}
\Psi_A^2 \xi^2 c^2 = [\Psi_A^2 \xi^2 c^2]_0(R)+\alpha [\Psi_A^2 \xi^2 c^2]_1(R)
\,,
\end{equation} 
and using the expressions of the pressure and the Mach number, we get an equation of the form,
\begin{equation}
\frac{d[\Psi_A^2 \xi^2 c^2]_0}{dR}+\alpha \frac{d [\Psi_A^2 \xi^2 c^2]_1}{dR}
=B_\star^2 M^2 \frac{d \Pi}{dR} \left[1+(\kappa+2m_1)\alpha\right]
\end{equation}
We see, like in the classical case, that the second term of the pressure is proportional to the first one such that, 
\begin{equation}
\frac{d [\Psi_A^2 \xi^2 c^2]_1}{dR} - (\kappa+2m_1)\frac{d[\Psi_A^2 \xi^2 c^2]_0}{dR}
=0 
\,.
\end{equation}
We deduce from the previous equation that the quantity $\epsilon$ defined by,
\begin{equation}
\epsilon(R)B^2_\star=[\Psi_A^2 \xi^2 c^2]_1- (\kappa+2m_1)[\Psi_A^2 \xi^2 c^2]_0={\rm cst.}
\,,
\end{equation}
is a dimensionless constant for all the field lines. 
To give explicitly $[\Psi_A^2 \xi^2 c^2]_0$ and $[\Psi_A^2 \xi^2 c^2]_1$, it may be useful to write,
\begin{equation*}
\Psi_A^2\xi^2c^2=\Psi_A^2\frac{(h\gamma\xi c)^2}{h^2}\left(1-\frac{(V^{\hat{\phi}})^2}{c^2}\right)-\frac{M_{Alf}^4 B_{\rm p}^2}{h^2}
\,.
\end{equation*}
Finally the calculation leads to,
\begin{eqnarray}
\label{epstt}
\epsilon&=&\frac{M^4}{h_z^2 h_*^4G^2(R^2+l^2)}\left[\frac{h_z^2 F^2(R^2+l^2)}{4R^2}-\frac{R^2}{(R^2+l^2)}\right.\\ \nonumber
&-& \left.(\kappa-2m_1)\frac{(R^2+l^2)}{G^2}\right]
- \frac{\nu^2(2e_1-2m_1+\delta-\kappa)R}{h_z^2(R^2+l^2)}\\ \nonumber
 &-&\frac{\nu^2l^2RG^2}{h_z^2(R^2+l^2)^3}+\frac{2\lambda^2}{h_z^2}\left(\frac{\Lambda^2N_B}{D} +\frac{\overline{\omega}_z}
 {\lambda}\right) + \lambda^2\left(\frac{\Lambda N_V}{h_* G D}\right)^2
\,.
\end{eqnarray}

This equation is similar to Eq. (71) in \cite{Melianietal06} and can be interpreted the same way. The parameter $\epsilon$ 
measures the efficiency of the magnetic rotator to collimate the flow. At the outflow base, $\epsilon$ is the relative difference of the transverse variation of internal energy
that is simply the exchange of work done by the macroscopic forces. As this is perpendicular to the flow axis,  this means that $\epsilon$ really measures the transverse 
force which collimates the flow and mainly its magnetic component.

Note that the quantity $-2m_1$ appears twice in Eq. (\ref{epstt}). First, it is 
associated with  $\kappa$, having a similar effect to the non spherically symmetric pressure in the term which is in factor of $M^4$.
Second, it is  associated with $2e_1$ in the term corresponding to the excess or the deficit of the gravitational energy not 
compensated by the thermal driving at the base of the jet.

To conclude we can also derive the magnetic collimation efficiency in a different form. After some calculations, we can put it in the following form,
\begin{eqnarray}
\epsilon&=&-\frac{\nu^2 h_\star^4}{\mu\gamma_z^2h_z^2}\frac{\partial}{\partial \alpha}\ln
\left. \left(\frac{P-P_0}{\rho_0\xi}\right)\right|_{\alpha=0}\\
&=&h_\star^2\frac{\nu^2}{\mu}\left(1-\frac{\mu}{\nu^2}\right)\frac{\xi_z^2}{\xi_\star^2}\left(\left.\frac{\partial}{\partial \alpha}\ln(\rho_0 \xi)\right|_{\alpha=0}-\kappa\right).\nonumber
\label{epsialf}
\end{eqnarray}

This new relation brings a link between the total enthalpy on the axis and its logarithmic variation with $\alpha$. 
In particular, the sign of $\epsilon$ seems to connect the balance between logarithmic variation of total enthalpy per unit of volume and the meridional increase of the pressure.
The factor indicates that $|\epsilon|$ probably tends to decrease for solutions which reach ultra-relativistic speed.

\section{Methodology for obtaining solutions}

\subsection{Numeral integration}

In Appendix C are given the coupled ordinary differential equations (\ref{FS}, \ref{Pressure}, \ref{G-F}) for the four  quantities of the Alfv\'en number, dimensionless 
radius, expansion factor and pressure,  $(M^2, G^2, F, \  \Pi)$.  
Details on the method and numerical techniques for the integration of these differential equations of the model system  can be 
found in \cite{ST94} and \cite{Melianietal06}. In brief, by using a Runge-Kutta scheme we start integrating from the Alfv\'en surface 
using the continuity relations. Integrating upwind, by adjusting the value of the slope of the derivative of the Alfv\'en number at the Alfv\'en transition, 
the unique 
value of $F_\star$ can be found that allows the crossing of the modified slow magnetosonic surface, for a given value of the pressure  at 
the Alfv\'en transition $\Pi_\star$. Then, integrating downwind 
 the value of $\Pi_\star$ can be further adjusted. The program  finds automatically, after several iterations of upstream and downstream 
integrations, the solution that crosses all critical points, a proxy that the asymptotic pressure converges towards the demanded value. In  
particular, in the following sections we have selected only solutions with the minimum possible value of $\Pi_\star$, the so-called 
limiting solutions (see \citealt{Sautyetal04}). We either have a collimated jet where at infinity $\Pi_\infty$ is minimum, or a conical wind  
where at infinity $\Pi_\infty$ is zero. If necessary, we can always add a constant value to the pressure $P_o$, to ensure that the pressure is positive everywhere 
in the flow.  

In the following,  we only outline briefly the points which differ from previous related studies for building our model in the framework of general relativistic  
magnetohydrodynamics in a Kerr metric. Some interesting spine jet solutions, both for their properties close to the black hole and at 
large distance, are presented in the next section \ref{Kerr}.  The presented solutions depend on a number of parameters and a systematic parametric study 
of the model is postponed for a following paper.

\subsection{Alfv\'en regularity conditions}

The regularity conditions at the Alfv\'en transition ($R=1$) for the azimuthal components and enthalpy have been already discussed 
in Sec. \ref{Toroidalfields}. In order that the field lines do not have a kink at the Alfv\'en transition (i.e., F is a continuous function across $R=1$),  we impose an extra regularity condition on the transfield equation giving the four physical quantities ($M^2, G^2, F, \Pi$)  at $R=1$ (\citealt{TT91}, \citealt{ST94}). 
In other words, similarly to the classical model \cite{ST94}, we should take appropriately into account this regularity condition at $R=1$, which gives the  ratio $\tau$ 
which is involved in the magnetic toroidal component,
\begin{eqnarray}
\tau&=&\displaystyle{\frac{N_B}{D}\Bigr\vert_*}=\displaystyle\frac {\displaystyle{
\frac{d N_B}{dR}\Bigr\vert_*}} {\displaystyle{\frac{d D}{dR}\Bigr\vert_*}} \nonumber\\
&=&\displaystyle{
\frac{h_*^2\left( \displaystyle{\frac{2}{1+l^2}}-F_*\right)-\displaystyle{ 
\frac{\mu(1-l^2)}{(1+l^2)^2}} -\displaystyle{\frac{h_* l \sqrt{\mu}\nu}{\lambda} \frac{l^2-3}{(1+l^2)^3}}}{p-\displaystyle{\frac{\mu(1-l^2)}
{(1+l^2)^2}}}}\,,
\end{eqnarray}
where p is the slope of the square of the Alfv\'en number at the Alfv\'en transition,
\begin{equation}
p\equiv\displaystyle{\frac{d M^2}{dR}\Bigr\vert_*}
\,.
\end{equation}

More specifically, the expansion factor F is determined by Eqs. (\ref{FS}) and (\ref{NF}).  And, in Eq. (\ref{NF}) appear the ratios of $N_V$ and  $N_B$ with $D$.  
Furthermore, $D$ appears in the denominator with a power higher by one than the powers of $N_V$ and  $N_B$ in the numerator.  Hence, 
in order that dF/dR does not diverge and the slope of $F(R)$ is continuous across R=1 we require that $\mathcal{N}_F  \cdot {D}\mid_*  = 0$ such that 
dF/dR is finite at $R=1$ wherein we have 0/0.
 Thus, near the Alfv\'en transition,  we may define the function $\mathcal{N}_F  \cdot {D} \sim 
\displaystyle{\mathcal{P}(M^2,G^2,F,\Pi)}$.  Then, in order to avoid a singularity at the Alfv\'en transition, a necessary condition is to 
choose $p,F_\star,\Pi_\star$ such that this function is zero, i.e.,  $\mathcal{P}(M^2=h_*^2,G^2=1,F_*,\Pi_*)=0$, and thus  
$\mathcal{N}_F \cdot D\mid_*=0 $. 
After some algebra we finally get a second degree polynomial for $F_\star$, namely, 
\begin{equation}
\mathcal{A}(p)F_*^2+\mathcal{B}(p)F_*+\mathcal{C}(p,\Pi_*)=0
\,,
\end{equation} 
with
\begin{eqnarray}
\mathcal{A}(p)&=&\lambda^2h_*^4+\frac{h_*^2}{4}\left(p-\frac{\mu(1-l^2)}{(1+l^2)^2}\right)^2\\
\mathcal{B}(p)&=&\left[\frac{1}{2}\left(\frac{\mu(1-l^2)}{(1+l^2)^2}-p\right)^3\right.\\ \nonumber
&-& 2\lambda^2h_*^2\left( p+\frac{2}{1+l^2}-\frac{2\mu(1-l^2)}{ (1+l^2)^2} \right.\\ \nonumber
&+& \left.\left.\frac{l \sqrt{\mu}\nu h_*}{\lambda}\frac{3-l^2}{(1+l^2)^3} \right)\right]\\
\mathcal{C}(p,\Pi_*)&=&\lambda^2 \left(p+\frac{2h_*^2}{1+l^2}-\frac{2\mu(1-l^2)}{(1+l^2)^2}\right. \\ \nonumber
&+& \left.\frac{l \sqrt{\mu}\nu h_*}{\lambda}\frac{3-l^2}{(1+l^2)^3}\right)^2\\ \nonumber
&+&\left(\kappa \Pi_*-\frac{1}{(1+l^2)^2}-\frac{2\lambda^2\mu}{\nu^2} \right.\\ \nonumber
&-&\left. 2\lambda^2-\frac{l^2(2\mu+\nu^2)}{(1+l^2)^3}\right) \left(p-\frac{\mu(1-l^2)}{(1+l^2)^2}\right)^2\,.
\end{eqnarray}
In this way, the regularity condition at the Alfv\'en transition is automatically satisfied and no more constraints are needed at $R=1$. 

\subsection{Effect of a nonspherical Alfv\'en number in a Schwarzschild metric}

To illustrate our model, we present in the following  two solutions, which are built in the framework of the Schwarzschild metric. 
The first one corresponds to a solution of a model presented in \cite{Melianietal06}, in which $m_1 = 0$. 

The chosen values of the other parameters are, 
$\lambda = 1.0, \, \kappa = 0.2, \, \delta = 1.2, \, \nu = 0.8, \, \ell = 0, \, \mu =  0.1, \, e_1 = 0$.   
We will compare this solution to a solution with the same parameters but by keeping the 
value of $m_1$ given by Eq. (\ref{m1alfv}), $m_1 = - 0.078$. 
In both solutions the value of $\Pi_\star$ is the minimum value of the limiting solution. Such solutions have the minimum amplitude 
of oscillations in the jet.

\begin{figure}[h]	
\centering
\includegraphics[width=7.5cm]{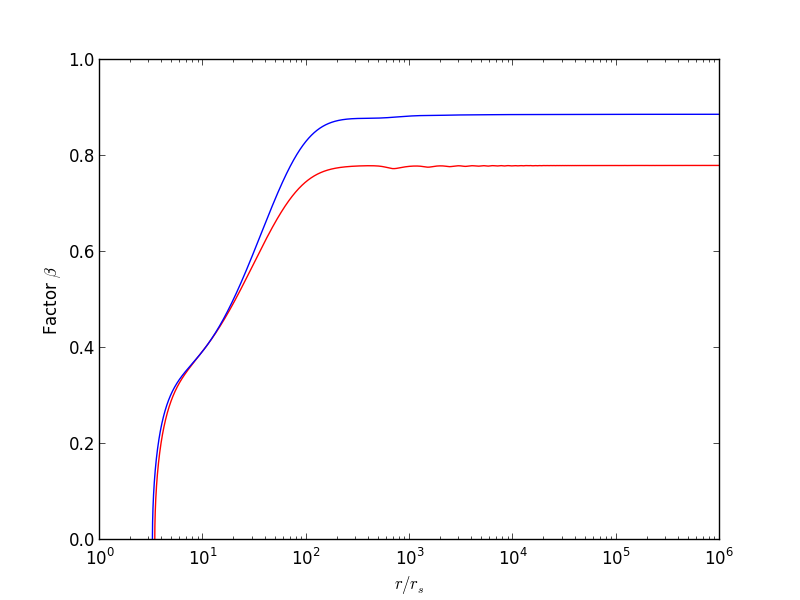}
\caption{Evolution of the radial velocity along the polar axis for solutions in a Schwarzschild metric, with $m_1=0$ ( blue) and 
$m_1 =-0.078$ (red). 
The second case has a smaller terminal velocity.
}   	
\label{FACB}
\end{figure}  

\begin{figure}[h]	
\centering
\includegraphics[width=9.5cm]{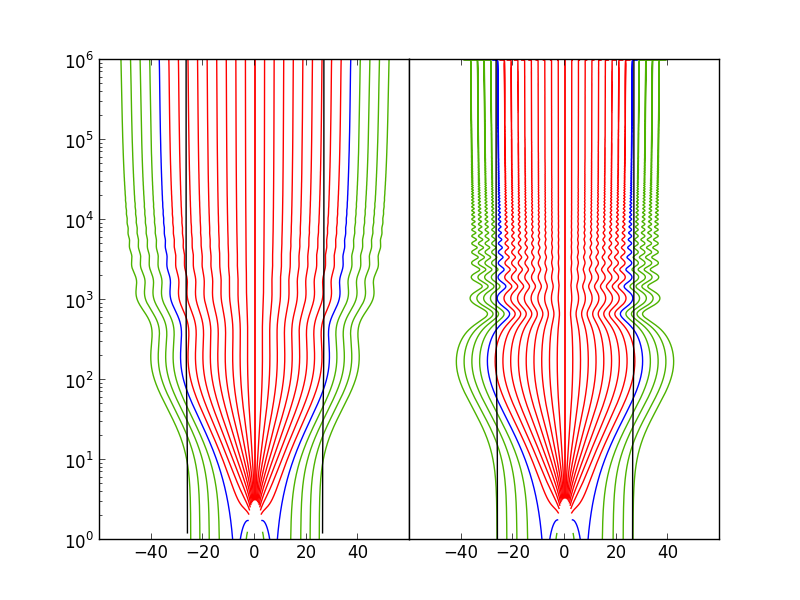}
\caption{Field lines for a solution in a Schwarzschild metric, with parameters 
$\lambda = 1.0, \, \kappa = 0.2, \, \delta = 1.2, \, \nu = 0.8, \, \ell = 0, \, \mu =  0.1, \, e_1 = 0$ 
and $m_1 = 0$ (left) and $m_1 = -0.078$  (right). Note that the case $m_1 = -0.078$ corresponds to a 
more tightly collimated jet. Lengths are in units of the Schwarzschild radius. The red lines are connected to the magnetosphere of the central object while 
the green lines are connected to the disk. The separating  line is in blue and the light cylinder in black.}   	
\label{NLC}
\end{figure}

In Fig. \ref{FACB}  the radial velocity on the polar axis is compared for the two solutions, while field lines in the poloidal plane 
are plotted in Fig. \ref{NLC}. 
Note that in Eqs.  (\ref{FS}) and (\ref{NM}) and Eqs. (\ref{FS}) and (\ref{NF}), giving the plasma acceleration and the variation of the expansion factor with the 
radius $R$, respectively, there are several terms proportional to the factor $(\kappa-2m_1)$.   
As $m_1$ is always negative, it is evident that $-m_1$ effectively  increases the transverse pressure gradient, which is proportional to 
$\kappa$. In other words, the effect of $-m_1$ is similar to the effect of $\kappa$ which enforces collimation for $\kappa > 0$.
Hence, taking into account a nonspherical Alfv\'en number ($m_1\neq 0$) introduces an extra collimation force which explains why the second solution with a non zero
$m_1$ is more collimated. Indeed, the width of the jet at infinity over its value at the base, ${G_{\infty}}/{G_0}$, decreases 
from 13.74 to 9.42, which is  similar to an increase of 
$\kappa$ and this fact can be checked directly by looking at the poloidal field lines shape  shown in Fig. (\ref{NLC}).

Since more tightly collimated solutions have a smaller super-Alfv\'enic acceleration, the second solution reaches a 
lower Lorentz factor asymptotically. 
Thus, similarly to $\kappa$, the introduction of a negative $m_1\neq 0$ leads to a decrease of the velocity because of the tighter collimation. 
The higher collimation reduces the pressure gradient along the axis, which in turn decreases the acceleration due to pressure driving on large distances (see \citealt{Sautyetal04}). 

Additionally, $m_1$ appears within the term $(\kappa+2m_1-\delta -2e_1)$ which appears in the plasma acceleration function $\mathcal{N}_{M^2}$ in Eqs.  
(\ref{FS}) and (\ref{NM}) and the function $\mathcal{N}_{F}$ determining the expansion factor ${F}$ in Eqs. (\ref{FS}) and (\ref{NF}). 
The first three terms $(\kappa+2m_1-\delta)$ arise from the variation across the field lines of the heat content $P/ \rho$ with $\alpha$, 
i.e., $\partial/\partial \alpha \{ [\Pi (R)M^2(R)] (1+\kappa \alpha )(1+m_1 \alpha )^2/(1+\delta \alpha] \} \simeq [ \Pi (R)M^2(R)] (\kappa + 2m_1 - \delta)$, 
while the fourth term $e_1$ is proportional to the variation of the total energy $\cal{E}$ with $\alpha$.  
The bigger this term is, the larger is the initial acceleration (see 
\citealt{ST94}), because it is linked with the distribution of the heating which opposes gravity to accelerate the outflow. As the weight of the plasma decreases 
with the latitude, then the pressure gradient increases along the axis resulting to a larger acceleration close to the base, as explained in \cite{TS92a}.
This term decreases rapidly as the Alfv\'en surface is reached.  
Thus, it is responsible only for the initial acceleration. With the parameter $m_1$ being negative, the second solution is more accelerated 
between the base and the Alfv\'en surface. The velocity of the second solution reaches the velocity of the first one at the 
Alfv\'en surface. This effect disappears far from the source. 

\section{Solutions in a Kerr metric}\label{Kerr}

In the following, we discuss four different solutions in a Kerr metric to illustrate the present model. A more detailed parametric study is 
postponed to a following paper. For the purposes of the present paper, we show three cylindrically collimated solutions 
with high asymptotic Lorentz factor, typical of AGNs and GRBs. Those solutions cross the "light cylinder" and are sorted with increasing
magnetic collimation efficiency parameter $\epsilon$. We also exhibit a conical solution crossing the "light cylinder" 
with high Lorentz factor and strongly negative $\epsilon$, something that was not possible with the previous relativistic meridionally self similar solutions. 

In order to get a Lorentz factor as high as possible in the asymptotic part of the collimated part of the jet, we know from the study of the 
classical solutions that among all cylindrical solutions, the limiting solutions with the lowest  value of $\Pi_{\star}$ reach the highest terminal 
velocity. These solutions are the so called limiting solutions in \cite{Sautyetal02}. As $\Pi_{\infty}$ is negative for the limiting solutions, we have
 to add a positive $P_0$ value to the pressure. Of course, it is always possible for  those cylindrical solutions to have a higher pressure $P_0$, 
  but by doing so it also increases the effective temperature, in particular in the asymptotic part. 
For the same set of parameters, it is also possible to get cylindrical solutions by increasing $\Pi_\star$. However, such solutions usually 
have a strong initial decollimation associated with a peak in the Lorentz factor and in temperature while the asymptotic jet is decelerated 
to lower Lorentz factors and smaller radii, a result we used to interpret the FRI/FRII dichotomy [cf. \cite{Melianietal10}]. 

\begin{table}[!h]
\begin{center}
\begin{tabular}{|*{7}{c|}}
\hline
  & $\boldsymbol{\lambda}$ & $\boldsymbol{\kappa}$ & $\boldsymbol{\delta}$ & $\boldsymbol{\nu}$ & $\boldsymbol{\mu}$  & $\mathbf{l}$  \\
\hline
$\mathbf{K1}$&1.0 & 0.2 & 2.3 & 0.9  & 0.1 & 0.05 \\
\hline
$\mathbf{K2}$&1.0  & 0.2  & 1.35  & 0.46223 & 0.1 & 0.05 \\
\hline
$\mathbf{K3}$&1.2  & 0.005  & 2.3  & 0.42 & 0.08 & 0.024 \\
\hline
$\mathbf{K4}$&0.0143  & 1.451  & 3.14  & 0.8 & 0.41 & 0.15 \\
\hline
\end{tabular}
\caption{Set of parameters used for the four selected solutions in the Kerr metric. $\mathbf{K1}$ is the solution displayed in Figs. \ref{TLC}, 
 \ref{OSC} and \ref{LOR} (blue line). Solution $\mathbf{K2}$ is displayed in Figs. \ref{LOR} (red line) and \ref{HIG}, while solution 
 $\mathbf{K3}$ is displayed in Figs.  \ref{ZON} and  \ref{TON}. Finally, solution $\mathbf{K4}$ is shown  in Figs.   \ref{CON} and \ref{LON}. 
}
\end{center}
\end{table}

\begin{table}[!h]
\begin{center}
\begin{tabular}{|*{5}{c|}}
    \hline
    &  $\boldsymbol{\epsilon}$ & $\mathbf{m_1}$ & $\boldsymbol{\Pi_{\star,lim}}$ & $\mathbf{r_0} / \mathbf{r_s}$ \\
   \hline
   $\mathbf{K1}$& -1.76& -0.062 & 0.826 & 5.72 \\
   \hline
   $\mathbf{K2}$& -0.04 & -0.234  & 0.216 & 1.57\\
    \hline
   $\mathbf{K3}$& 0.55 & -0.326 & 0.189 & 2.55\\
  \hline
   $\mathbf{K4}$& -5.84 & -0.004 & 0.255 & 1.39\\
   \hline
\end{tabular}
\caption{Output parameters for the four solutions in the Kerr metric. Those parameters result from the integration of the equations. }
\end{center}
\end{table}

Solutions $\mathbf{K1}$ and $\mathbf{K2}$ have been obtained for maximally rotating black holes, i.e. $a_H$ close to $1$ 
($a \simeq r_s/2$). 
In solution $\mathbf{K4}$ the value of $a_H$ has been fixed to $0.73$  ($a=0.73 r_s/2$). We do not expect all black holes to be maximally rotating. For instance in M87, 
the dimensionless spin should be above $0.65$ (i.e. $a> 0.65 r_s/2$), (\citealt{Lietal09}) but not too close to one. 
Other examples can be found 
and we will use for $\mathbf{K3}$ the value $a_H= 0.6$ ($a=0.6 r_s/2$)  adopted in \cite{Mertens16} for M87.

\subsection{A mildly relativistic  collimated solution with oscillations (K1)}

\label{RecOscSolK}

The collimated solution $\mathbf{K1}$ corresponds to an over-pressured outflow ($\kappa \geq 0$) in a Kerr metric. As  
$\epsilon \leq 0$, the collimation of the jet is not fully magnetic but it has a significant contribution by the gas pressure, at least during the 
phase of strong acceleration up to $\simeq 30 r_s$.  
In this solution, field lines are strongly oscillating compared to the two previous solutions in the Schwarzschild metric. The outflow undergoes 
a series of strong oscillations connected to the balance between the toroidal magnetic tension and the decollimation forces (centrifugal and 
electric forces and transverse pressure gradient) of the plasma. 

\begin{figure}[h]	
\centering
\includegraphics[width=9.0cm]{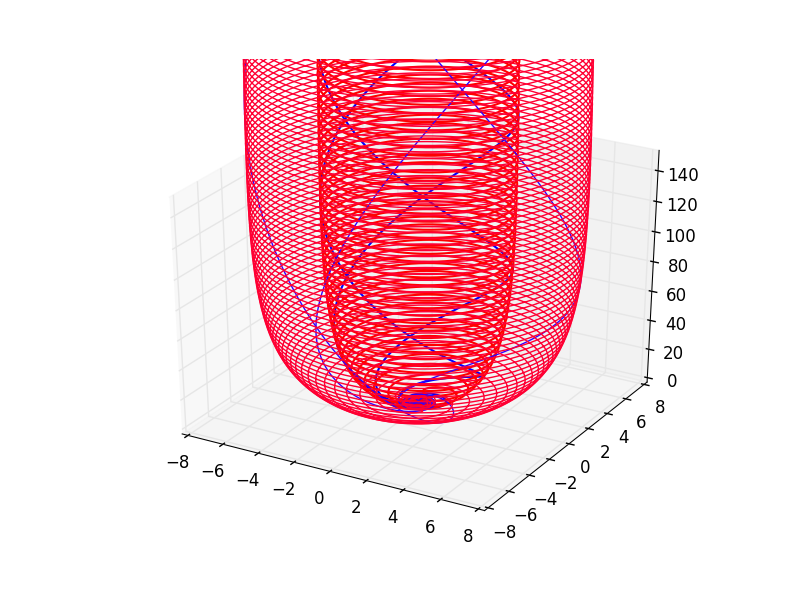}
\caption{3D representation of the field lines and streamlines for the thermally collimated solution $\mathbf{K1}$ at the base of the jet and for two flux tubes. 
The blue lines correspond to streamlines, the red lines to magnetic field lines. The length is in units of the Alfv\'en radius, i.e., ten times the 
Schwarzschild radius.}   	
\label{TLC}
\end{figure}
\begin{figure}[h!]
\centering
\includegraphics[width=9.5cm]{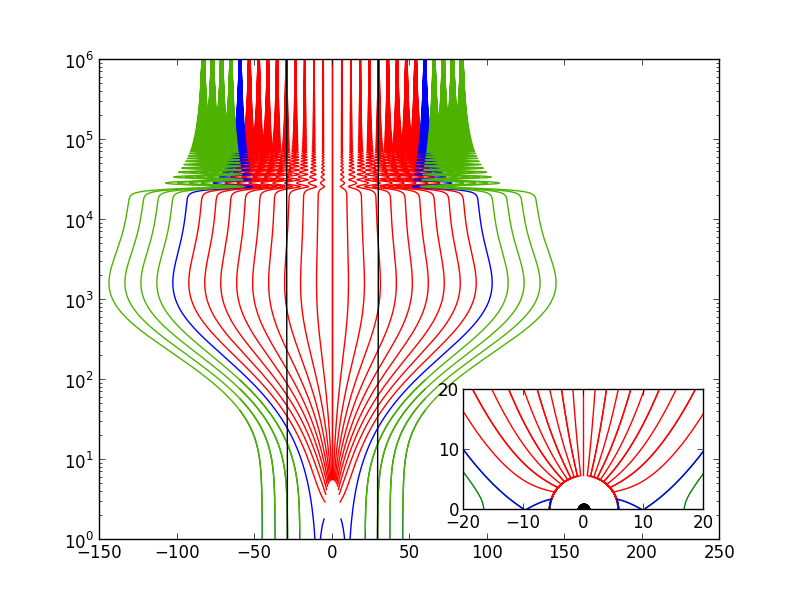}
\caption{Poloidal field lines and "light cylinder" for the thermally collimated solution $\mathbf{K1}$, for $\lambda = 1.0, \, \kappa = 0.2, \,
 \delta = 2.3, \, \nu = 0.9, \, \mu =  0.1, \, \ell = 0.05, \, e_1 = 0$. 
The length unit is the Schwarzschild radius. 
}
\label{OSC}
\end{figure}

The significant contribution of the transverse pressure gradient also explains these strong oscillations in the
flow as it is also shown in the classical solutions. The  parameters of this solution are displayed in the first row 
$\mathbf{K1}$  of  Table 1 and the output values of  $m_1 $ and $\epsilon$ in Table 2. 
 Note that $m_1 = -0.062$ is a relatively small value, which clearly indicates that the Alfv\'en surface is almost spherically symmetric 
 in this case. As a consequence the light 
 cylinder is relatively far from the jet axis. Most of the central field lines (see the inner 5 to 7 central red lines in Fig. \ref{OSC}) remain within 
 the "light cylinder", which means that despite the important role of the magnetic field in the collimation, the jet is pressure 
 or enthalpy driven in the relativistic case. 
 However conversely to the classical solutions, the electric field is the dominant decollimating force for the lines that cross the 
 "light cylinder". 
 With this decollimation and expansion after the Alfv\'en surface, is associated a strong pressure
 gradient yielding a strong acceleration of the jet in the super-Alfv\'enic regime.  More details on this will be given in the next solution. 
The pressure gradient is the gas pressure gradient close to the axis but assisted by the toroidal magnetic pressure outside the "light cylinder". 
This is similar to superfast flows in radially self similar models for disk winds (see \citealt{VlahakisKonigl03a} and 
\citealt{VlahakisKonigl03b}).  Moreover, in the relativistic regime the inertia
 increases faster when the flow is accelerated such that the collimation from the magnetic field is delayed to larger distances. 
 
 In Fig. \ref{OSC} we see that the expansion after the 
 Alfv\'en surface is strong and leads to a late acceleration of the flow. 
After the large expansion, the jet recollimates smoothly and consequently decelerates slightly because of the compression.

In Fig. \ref{TLC},  we clearly see that there is a strong azimuthal magnetic field although the scale of the figure tends to exaggerate this 
phenomenon. 

The Lorentz factor of this solution reaches a relatively small value around $3.7$, typical of AGN jets, which are not
too powerful, like some of the FRI radio-galaxies (see Fig. \ref{LOR}).  

\subsection{A highly relativistic collimated solution with oscillations (K2)}

The solution $\mathbf{K2}$  is collimated and has an extremely high Lorentz factor, which may be typical of 
GRBs. 
This $\mathbf{K2}$ model corresponds to the values of the  parameters  given in the second line of Table 1, 
i.e.,  $\lambda = 1.0, \, \kappa = 0.2, \, \delta = 1.35, \, \nu = 0.46223, \, \ell = 0.5, \, \mu =  0.1, \, e_1 = 0$ 
and the second line of Table 2 for the output parameters,  $m_1 = -0.234$ and $\boldsymbol{\epsilon} =-0.04$. 

\begin{figure}[h]	
\centering
\includegraphics[width=8.5cm]{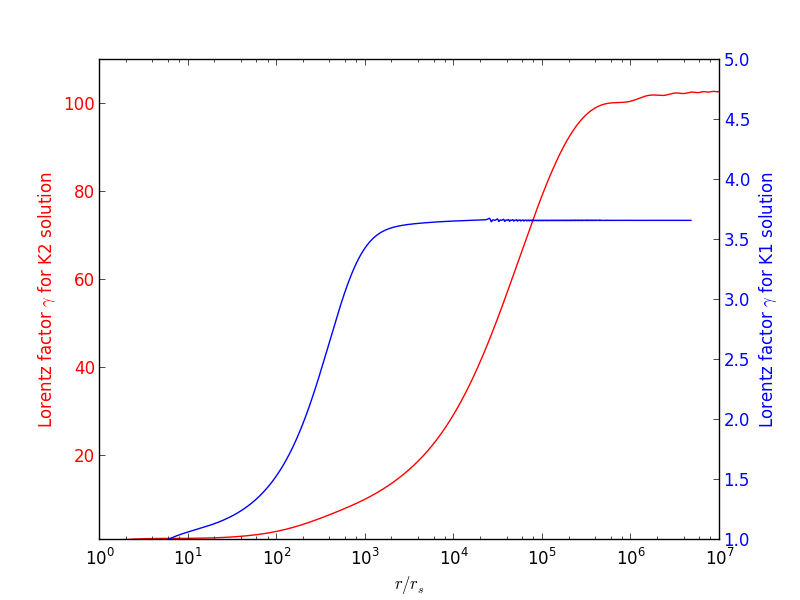}
\caption{Lorentz factor $\gamma$ for the $\mathbf{K1}$ (blue line) and $\mathbf{K2}$ (red line) solutions.
Distances are given in Schwarzschild radius units.}   	
\label{LOR}
\end{figure}

\begin{figure}[h]	
\centering
\includegraphics[width=9.5cm]{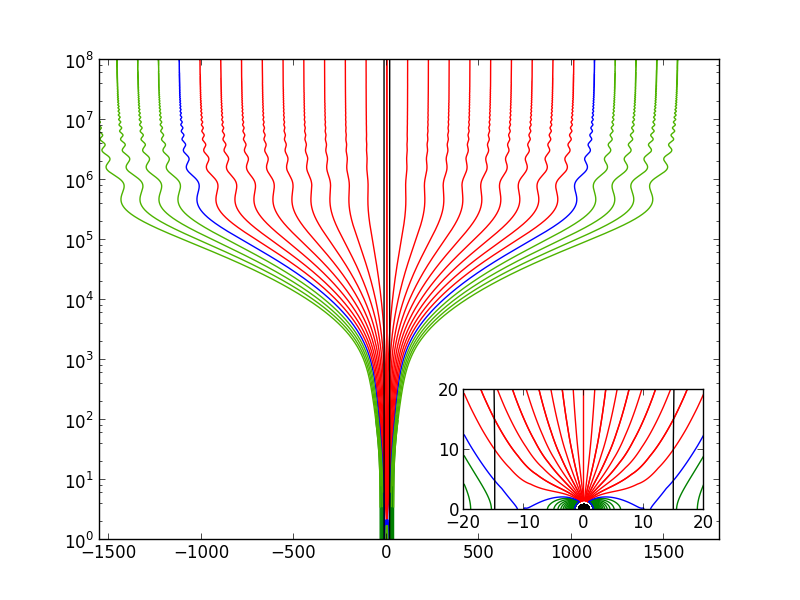}
\caption{Poloidal field lines and "light cylinder" for the $\mathbf{K2}$ solution, i.e. for  $\lambda = 1.0, \, \kappa = 0.2, \,
 \delta = 1.35, \, \nu = 0.46223, \, \mu =  0.1, \, \ell = 0.05, \, e_1 = 0$.
Distances are given in Schwarzschild radius units.}   	
\label{HIG}
\end{figure}

For this model, the outflow starts very close to the black hole horizon at $r_0=1.57\, r_s$, (see Fig. \ref{HIG}),  
and thus at the base of the jet the effects of general relativity play an important role. The final velocity is highly relativistic, as it is shown in Fig. \ref{LOR}.  

The parameter $\nu$ of the  $\mathbf{K2}$ solution is accurately adjusted (to the fifth digit), such as to obtain 
 a rather high Lorentz factor (larger than 100). 
 This proves the versatility of the model which handles any magnitude of Lorentz factors. In order to obtain such high Lorentz factors, we must 
 carefully tune  the parameter directly linked to gravitation, $\nu$, as mentioned above. 
 The same parameter is also responsible for the thermal  acceleration in the classical model (\citealt{ST94}). 

\begin{figure}[h]	
\centering
\includegraphics[width=9.5cm]{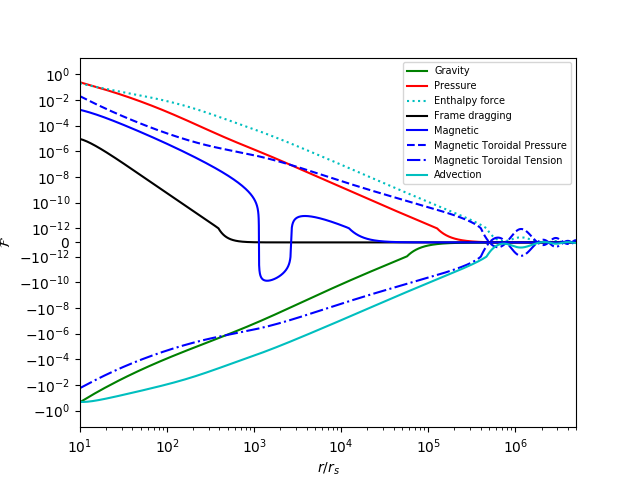}
\caption{Plot of the longitudinal forces, i.e. along the field line, for the $\mathbf{K2}$ solution, along the line 
$\alpha=0.01\alpha_{\rm lim}$. 
Distances are given in Schwarzschild radius units.}   	
\label{longitudinalforce}
\end{figure}

\begin{figure}[h]	
\centering
\includegraphics[width=9.5cm]{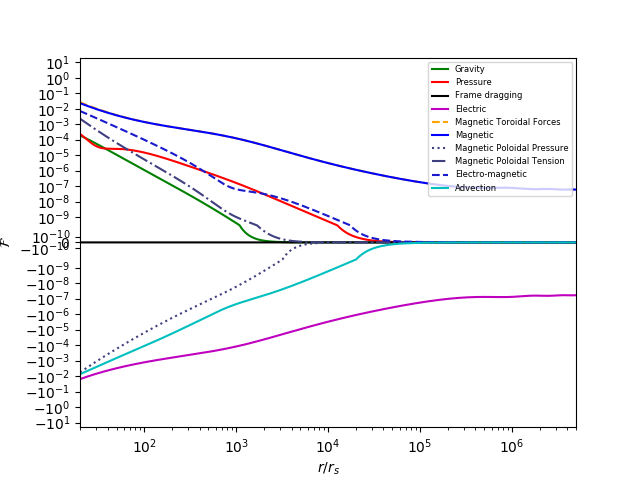}
\caption{Plot of the transverse forces, i.e. perpendicular to the field line, for the $\mathbf{K2}$ solution, along the line $\alpha=0.01\alpha_{\rm lim}$. 
We see that the Lorentz force is collimating and is balanced by the electric force that decollimates. Distances are given in Schwarzschild radius units.}   	
\label{transverseforce}
\end{figure}

In Figs. \ref{longitudinalforce} and \ref{transverseforce} we plot the forces, along and perpendicular to a field line defined by $\alpha=0.01\alpha_{\rm lim}$ where $\alpha_{\rm lim}$ is the dimensionless magnetic flux between the inner jet and an external accretion disk wind.

The strong decollimation associated with the slow acceleration enhances the electric force as in the previous solution $\mathbf{K1}$. 
This can be seen in Fig. \ref{transverseforce}. However due to the higher rotation here, more field lines cross the light cylinder, which is very close
to the axis, such that the decollimation from the electric field is much stronger in this solution. 

The feedback of this strong electric field is to increase further the decollimation beyond the Alfv\'en surface at large distances. Again, the large  expansion increases the pressure and enthalpy gradient 
as seen in Fig. \ref{longitudinalforce}. Thus, the pressure force increases, resulting to  a 
very long acceleration phase up to $10^6 r_s$. Thus, the thermal acceleration becomes very efficient and the plasma reaches asymptotically an extremely high 
Lorentz factor. Indeed, the jet radius increase of this solution is also very high with an expansion factor ${G_{\infty}}/{G_0} \simeq 1300$, as it can be  
seen in Fig. \ref{HIG}.

The value of $\epsilon$ is still negative but very close to zero ($\epsilon = -0.04$). As in the classical case, this means the magnetic 
efficiency to collimate the flow is higher in this model at larges distances. However the decollimating force that ensures the equilibrium is no longer the 
centrifugal force or the pressure gradient but the electric force on the lines that cross the "light cylinder". This is a specific feature of 
relativistic jets.

\begin{figure}[h]	
\centering
\includegraphics[width=8.5cm]{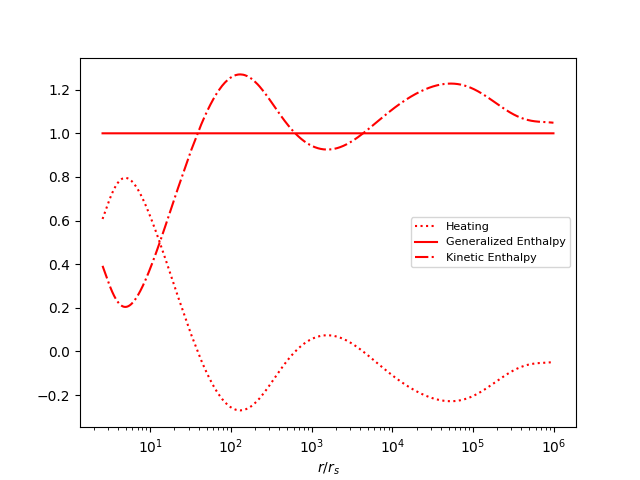}
\caption{Relative normalized contribution to the total energy of the kinetic enthalpy $h \gamma \xi_k$ and the external heating distribution $h \gamma Q/c^2$ along the axis.}   	
\label{ERA}	
\end{figure}

\begin{figure}[h]	
\centering
\includegraphics[width=8.5cm]{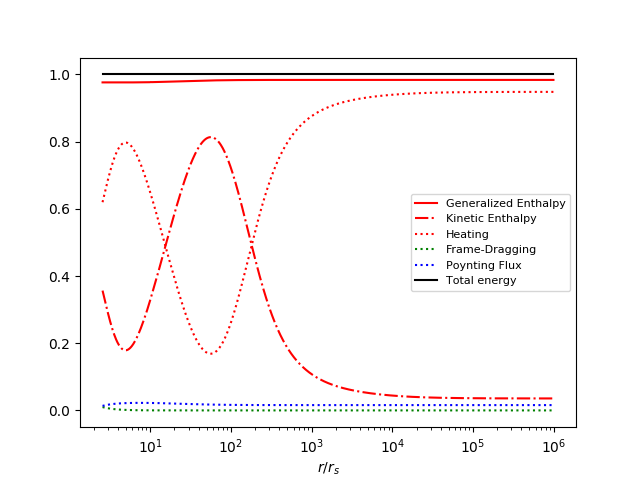}
\caption{Relative normalized contribution to the total energy of the kinetic enthalpy $h \gamma \xi_k$ and the external heating distribution $h \gamma Q/c^2$ along the streamline $\alpha = 0.05 \alpha_{\rm lim}$}   	
\label{ERL}	
\end{figure}

To analyze further the jet acceleration we may calculate the contribution of the different components of the total energy and the conversion of the magnitude 
of each component to another form along the streamlines. 
First, we want to get a physically acceptable heating term which goes to zero at infinity along the axis of the flow. 
By defining the enthalpy in Eqs. (\ref{TETENT}) and (\ref{ENTKIN}) and its analytical
expression in our model in Eq. (\ref{INV6}),  we may fix a streamline limiting the zone where the pressure is positive. 
Once this field line has been chosen, e.g. at $\alpha = 0.4 \alpha_{\rm lim}$, the dimensionless pressure ${P_0}/{\rho_\star c^2}$ can be calculated. 
In the particular case presented here we have taken ${P_0}/{\rho_\star c^2}=4.3 \cdot 10^{-7}$ . 
The pressure ${P_0}$ in Eq. (\ref{Pdef}) is chosen such that the gas pressure is equal to zero, when $\Pi(R)$ reaches its minimum value. 
For some value of $\alpha$, the term $P_0 = P_0(\alpha)=- [{B_\star^2}/{8\pi}]\Pi_{\rm min}(1+\kappa \alpha)$ 
will be too large to insure that $Q$ goes to zero at infinity, along the axis. 
Indeed, $\alpha$  can take any values below a maximum $\alpha_{\rm max}$.
We can consider that the solution can be valid only in the region defined wherein the pressure is positive. 
When $\alpha$ is fixed and ${P_0}/{\rho_\star c^2}$ is deduced, we are able to calculate $\xi_\star$. In this particular case, 
a value $\xi_\star \simeq 78$ is chosen.  

Fig. \ref{ERA} shows the normalized total energy on the axis $\mathcal{E}/c^2 = h \gamma \xi$, the kinetic component $h \gamma \xi_K$ and the external 
heating $h\gamma Q/c^2$. 
We find a decrease of the external heating and a related increase of the kinetic part. Thus,  the kinetic enthalpy represents the major component up to $r=10^2 r_s$.

Fig. \ref{ERL} shows the same energetic distribution, but on a streamline with $\alpha=0.05\alpha_{\rm lim}$. Out of the polar axis, there are extra energetic components, 
such as the frame-dragging and the Poynting fluxes. Both these energetic contributions are very small on this field line, as compared to the total energy. Hence,  the jet is enthalpy-driven from 
the axis right up to the limiting line. We also note that at the base of the jet, the frame-dragging energy is of the same order with the Poynting flux. Contrary to the energetic distribution along the axis, 
the external heating constitutes the larger part of energy at infinity.  
While along the axis a high value of $\gamma_\infty \simeq 100 $ is obtained, the Lorentz factor at infinity on this particular line is $\gamma_\infty\simeq 3.6$. 
Hence, since the acceleration of the plasma and the resulting final flow speed at infinity on this particular field line are small, 
the external heating is not consumed to accelerate the flow and therefore it is left unused at infinity, contrary to what happens along the axis, Fig. \ref{ERA}.

\subsection{A mildly relativistic collimated solution without oscillations (K3)}

\cite{Mertens16} recovered from VLBI imaging a detailed two-dimensional velocity field in the jet of M87 at sub-parsec
scales. They confirmed the stratification of the flow from the very beginning of the jet and identified a relativistic sheath,
i.e. an accelerating layer, which is launched from the inner part of the accretion disk at a cylindrical distance around 5 $r_s$.
\cite{Mertens16} interpret this outer sheath layer as the internal part of an external disk wind. They also
 interpret the inner spine jet as a component coming from the internal accretion disk.
However, it is not clear that the inner spine necessarily has to be  one of the disk wind components. 
Instead we propose that the inner spine jet originates from the black hole corona and 
has a higher Lorentz factor. The authors note that this fast inner spine jet cannot be  
detected in their data because of its lower emissivity compared to the one of the sheath layer or because its speed is too high. 

We propose here, as an alternative scenario, that the spine beam may originate from the magnetosphere of the black hole, 
either connected to the black hole itself as in the Blandford-Znajek-Penrose mechanism, or, connected to the inner part of the accretion disk. 
In the first case, the spine jet would be a leptonic plasma, with a hadronic one for the second alternative. 
In such a case, we can model the spine jet with our meridionally self-similar solutions. 
Note that conversely to Poynting flux dominated models, our model is valid on the jet axis.

The Lorentz factor profile inferred by \cite{Mertens16}  (see their Fig. 19) supposes that the velocity structure observed in the 43 GHz 
VLBA maps at a deprojected distance of $z \simeq 10 mas \simeq 1300 r_s$ is due to the sheath layer. 
The Lorentz factor of the sheath has a value $\gamma \simeq 2.4$ at $R\simeq 1000 r_s$.  

 From the curve deduced for the spine jet with an approximated MHD disk wind solution, those authors 
 following \cite{Andersonetal03}, model the acceleration and collimation of the flow with a Lorentz factor $\gamma \approx 7$. We 
 propose that it is possible to construct a MHD solution, along the lines of the model analysed in this paper, by choosing a positive value 
 of $\epsilon \approx 0.5$ and a similar velocity profile. The solution  $\mathbf{K3}$ we
 present here has a Lorentz factor $\gamma \simeq 5$ at $R\simeq 1000 r_s$. By appropriately tuning the parameters, we could obtain even higher Lorentz 
 factors from $7$ to $10$, as it is observed at the distance of  the knot HST-1, where this velocity is observed in the optical band. 
 However, the spine jet is deboosted relatively to the sheath and a precise measurement of its Lorentz factor is difficult.

Solution $\mathbf{K3}$ with a maximum Lorentz factor on the axis $\gamma \simeq 5.5$ has an asymptotic spine jet radius 
${G_{\infty}} \simeq 20 r_s$. At this distance from the axis, the Lorentz factor has dropped to a value  $\gamma \simeq 2.4$ consistent with the radius and the Lorentz factor 
at the inner observed distance of the outer sheath jet of \cite{Mertens16}. Thus, our solution may model the initial spine jet inside the sheath layer. 
However, to confirm that, we need to use this initial solution in simulations similar to those in \cite{Hervetetal17}. 
The spine jet/sheath jet interaction will probably produce shocks and rarefaction waves that may further accelerate the jet.

So, although we can obtain such a type of collimated solution for different sets of the parameters, we focus here on the 
specific solution $\mathbf{K3}$, where  $\lambda = 1.2, \, 
\kappa = 0.005, \, \delta = 2.3, \, \nu = 0.409, \, \ell = 0.024$ and  $\mu =  0.08$. Compared to the other Kerr solutions studied in 
this paper, $\lambda$ 
is higher and $\kappa$ is very small, leading to a solution with a positive value of the magnetic collimation efficiency, $\epsilon 
=0.55$.  Another advantage of this solution is the fact that the pressure depends only very weakly on the magnetic flux function, i.e. 
on a particular field line. 

Interestingly, in \cite{Mertens16},  the radius  of the base of this spine jet is equal to $r_0=2.4r_s$,  
which is consistent with their model of a disk wind solution, by fixing the jet shape and solving the Bernoulli equation. In our 
self-similar solution, we have a similar radius for the magnetospheric polar cup, where our jet solution starts. 
Clearly,  this is an alternative scenario.

Moreover, the angular velocity of the field lines anchored in this polar cup above the black hole can be calculated from our parameters,
\begin{eqnarray}
\Omega_\star \!  \!  \! \! \!&=&\! \! \! \!  \! \frac{c \mu}{r_s} \left[\frac{\sqrt{\mu} \lambda}{\nu}\sqrt{1-\frac{\mu}{(1+l^2)}}+\frac{l \mu}
{(1+l^2)^2}\right] \simeq 6.2 \times 10^{-2} \frac{c}{r_s}\,.
\end{eqnarray}
By taking the value of the M87 distance and the black hole mass as in \cite{Mertens16}, we can calculate $\Omega_\star$ in the context of this {\bf K3} solution. This 
value may directly be compared to the values they deduced in two jet regions from the conservation of total energy and angular 
momentum fluxes in the approximation of special relativity. 

Hence, we find  $\Omega_\star \simeq 1.03 \times 10^{-6} s^{-1}$, a value which is for the spine jet almost the same value
as the isorotation frequency of a Keplerian speed at the launching location of the sheath layer. 
It also corresponds to the initial toroidal velocity of the Blandford \& Payne (1982) mechanism. 

Our model corresponds to an alternative configuration, because the spine jet may either originate from the Keplerian disk, in which case it would be hadronic, 
or form via the generalized Penrose-Blandford-Znajek mechanism. In this second alternative the jet would be a leptonic beam with an angular 
frequency proportional to the spin of the central black hole. 
The Blandford-Znajek mechanism allows to extract energy from the black hole when $0\leq\Omega\leq\omega_{BH}$, with a maximum value for $0.5 {\omega_{BH}}$.
 Note that $\omega_{BH}$ is, by definition, the angular velocity of ZAMO at the location of the outer event horizon and is given by,  
 \begin{eqnarray}
\omega_{BH} &=& \frac{a_H c}{r_s \left(1 + \sqrt{1 - a_H^2}\right)}
\end{eqnarray}
where $a_H$ is the dimensionless spin of the black hole in units of the gravitational radius $r_s/2$. 
Indeed simulations of such Poynting-dominated and force-free jets  (\citealt{Tchek15}) have shown that the angular speed of a field line 
 anchored in the magnetosphere is about half the black hole angular speed $\omega_{BH}$. Our value of $\Omega_\star$ is one third of $0.5 \omega_{BH}$ (\citealt{Nathanail14}).
In order to determine if the spine jet of our solutions originates from a Keplerian disk or from a black hole via 
a generalized Penrose-Blandford-Znajek mechanism, we need to solve the MHD 
equations up stream up to the black hole horizon. Moreover, in order to model the full jet of M87, a complete MHD simulation including 
a disk wind and a spine jet has to be developed. Something that should be done in the future.

Note that as it is already known, at the interface of the spine jet and the sheath layer, a re-collimation shock may occur, producing  
compression and rarefaction waves which may accelerate the flow (\citealt{Hervetetal17}). For those reasons we have chosen here to adjust 
the value of $\nu$ to $\nu = 0.42$, in order to get the solution
$\mathbf{K3}$, with a lower Lorentz factor but suppressing completely the oscillations of the field lines. 
The value calculated above for $\Omega_\star$ is not changed, but radio emission maps, which are obtained for the M87 jet 
will be produced by re-collimation shocks due to the interaction between the fast spine jet and the sheath layer.

\begin{figure}[h]	
\centering
\includegraphics[width=8.5cm]{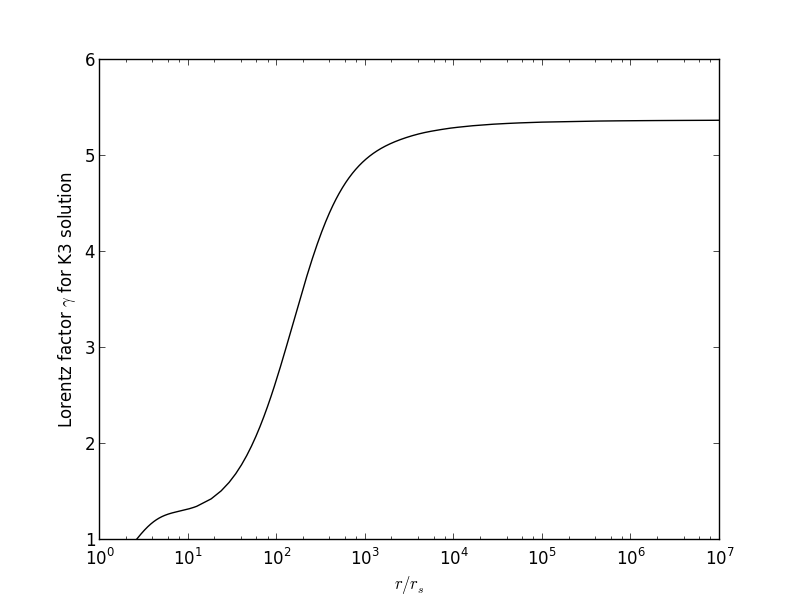}
\caption{Lorentz factor for the non-oscillating collimated solution $\mathbf{K3}$.}   	
\label{ZON}	
\end{figure}

As it can be seen in Fig. \ref{ZON}, for the $\mathbf{K3}$ model the Lorentz factor reaches a nearly constant value at the 
distance of the $B$ structure observed in the 43 GHz VLBA maps and $\gamma$ is larger than the one deduced for the sheath layer 
by \cite{Mertens16}.   
As explained above, the interaction between the sheath layer and the spine jet can induce a bulk flow acceleration. 
However, as we mention the Lorentz factor $\gamma$  is maximum along the axis and decreases with latitude such that at its 
outer boundary it matches the sheath layer value. 

\begin{figure}[h]	
\centering
\includegraphics[width=9.5cm]{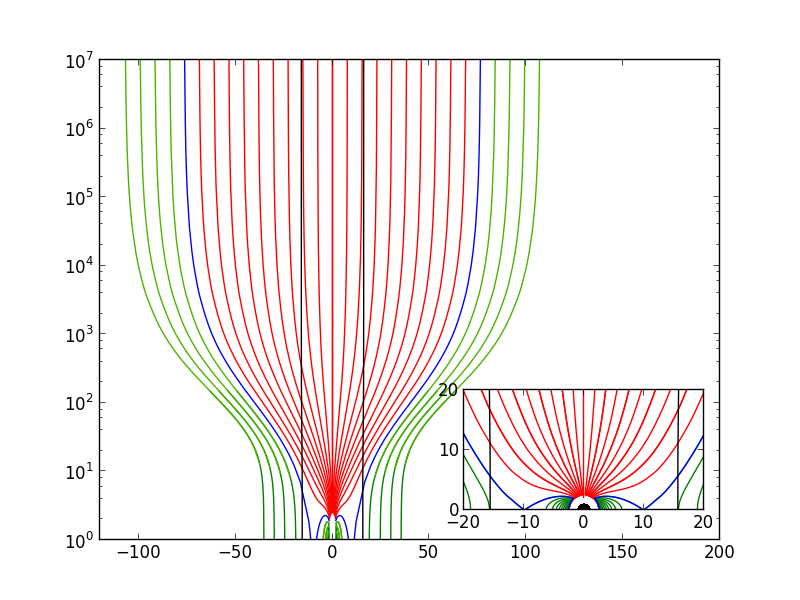}
\caption{Poloidal field lines and "light cylinder" for the non-oscillating collimated solution $\mathbf{K3}$, i.e. for  $\lambda = 1.2, \, \kappa = 0.005, \,
 \delta = 2.3, \, \nu = 0.42, \, \mu =  0.08, \, \ell = 0.024, \, e_1 = 0$. Distances are given in Schwarzschild radius units.}   	
\label{TON}
\end{figure}

The values of the parameters of the 
$\mathbf{K3}$ model are given in the third line of Tables 1 and  2. The radius at the base of this spine jet is equal to $r_0=2.55r_s$ and the magnetic efficiency 
to collimate the flow is larger compared to the other two Kerr solutions discussed in this paper, as $\epsilon$ is now positive, which means that the jet is fully magnetically collimated. The field lines displayed for $
\mathbf{K3}$ in Fig. \ref{TON}
are nearly cylindrical above the equatorial plane, at distances $10^4$ $r_s$, with a smooth flaring occuring after the Alfv\'en surface.

\subsection{Conical solution}

The parameters of the conical solution $\mathbf{K4}$ are given in Tables 1 and 2. Such radial solutions could be useful to describe 
relativistic non collimated outflows, such as those seen in association with radio quiet galaxies, like Seyfert  galaxies. Conical solutions could also be useful to 
model GRBs, wherein an unstable non collimated relativistic wind may fragment into small pieces under some instabilities. In such case the 
apparent collimation of the GRB would be due to fragmentation -- see \cite{MelianiKeppens10, vanEertenetal11}.

A conical solution is a solution in which the limiting value of $F$ at infinity is $0$. In this solution, the spherical part 
of the Mach number diverges, $M \rightarrow \infty$. The same effect occurs for the cylindrical radius of the flow, $G$. 
To get this conical solution, we started with the parameters used for modeling the Solar Wind, as in \cite{Sautyetal05}.  
The magnetic collimation parameter must be strongly negative, $\epsilon/(2\lambda^2) \leq 0$. 
We adjust the solution to the relativistic case and increase the velocity, by taking a larger value for $\delta$. 
For the radial solution $\mathbf{K4}$, the parameters $\lambda$ and $\kappa$ are
adjusted as well in order to get a terminal Lorentz factor larger than 8. Thus, we find 
a conical solution $\mathbf{K4}$ for the following set of parameters $\lambda = 0.0143, \, \kappa = 1.451, \, \delta = 3.14, \, 
\nu = 0.8, \, \ell = 0.15, \, \mu =  0.41, \, m_1 = -0.004$ and $ e_1 = 0$.

\begin{figure}[h]	
\centering
\includegraphics[width=9.5cm]{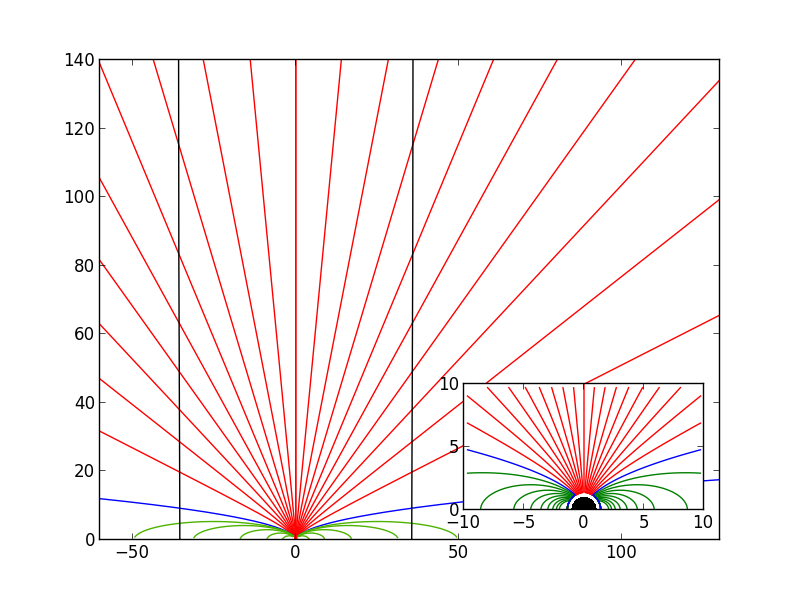}
\caption{Poloidal field lines and "light cylinder" for the conical solution $\mathbf{K4}$, i.e. for  $\lambda = 0.0143, \, \kappa = 1.451, \,
 \delta = 3.14, \, \nu = 0.8, \, \mu =  0.41, \, \ell = 0.15, \, e_1 = 0$.
Distances are given in Schwarzschild radius units.}   	
\label{CON}
\end{figure}

\begin{figure}[h]	
\centering
\includegraphics[width=8.5cm]{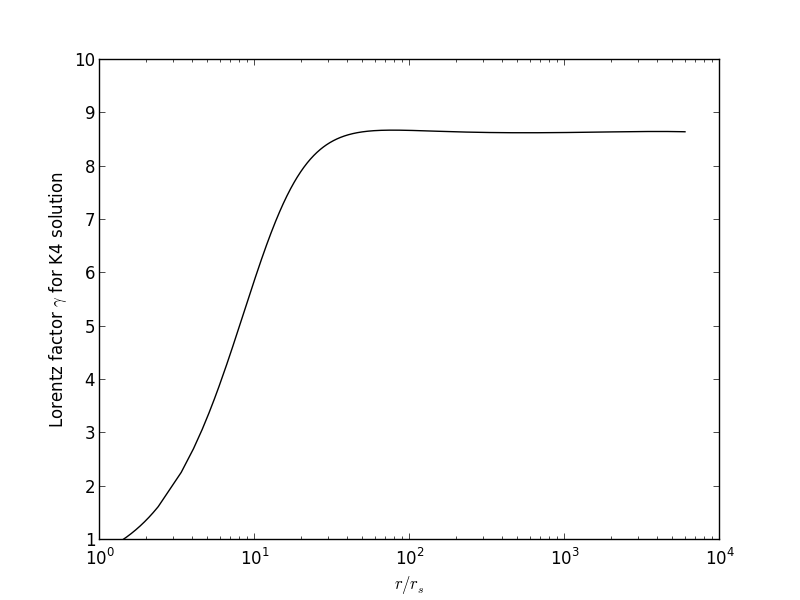}
\caption{Lorentz factor for the conical solution $\mathbf{K4}$.
Distances are given in Schwarzschild radius units.}   	
\label{LON}	
\end{figure}

For this set of parameters we get a much more negative value for the magnetic
collimation efficiency parameter, $\epsilon=-5.68$. The solution quickly reaches the conical regime 
(see Fig. \ref{CON}), and most of the field lines cross the "light cylinder" plotted as a black solid line.  
The axial radial velocity profile is plotted in Fig. \ref{LON}. As it can be  seen, high Lorentz factors are obtained.  

\subsection{Magnetic collimation efficiency versus black hole spin}

As we already mentioned, the constant $\epsilon$ is a measure of the efficiency of the 
magnetorotational forces to collimate the flow. 
We have studied how $\epsilon$ relates to the black hole spin $a_H=2 l $ / $ \mu$, 
keeping unchanged all the other parameters, $\lambda, \, \kappa, \, \delta, \, \nu, \, $ and $ \mu $.

\begin{figure}[h]	
\centering
\includegraphics[width=9.5cm]{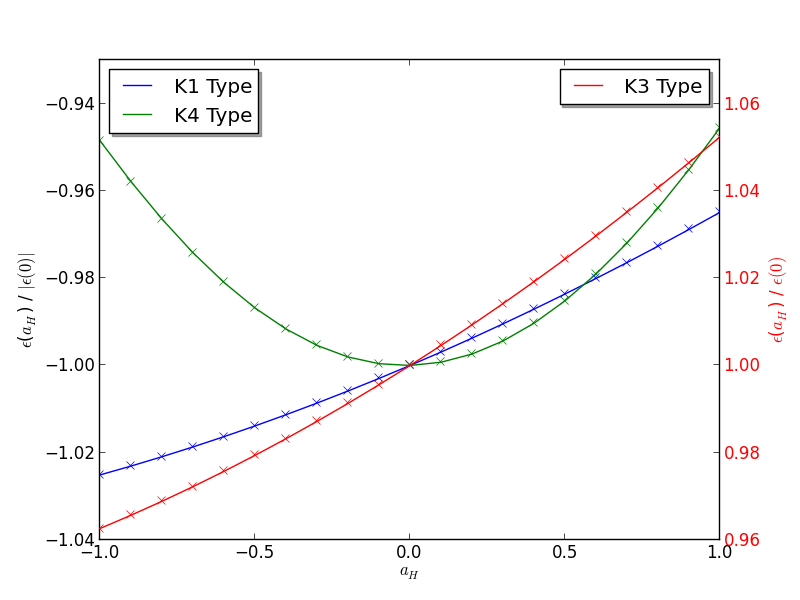}
\caption{Variation of the magnetic collimation efficiency parameter $\epsilon$ versus the black hole spin parameter $a_H$ for  
$\mathbf{K1}$-type, $\mathbf{K3}$-type and $\mathbf{K4}$-type solutions. The value of $\epsilon$ has been normalized by $| \epsilon(0) |$ when 
it is negative and by $ \epsilon(0) $ when it is positive.}   	
\label{EPL}
\end{figure}

In Fig. \ref{EPL}, we plot $\epsilon$ versus the black hole spin $a_H$, for several cases we studied in the 
context of our model. 
First we note that for solutions with parameters similar to the $\mathbf{K3}$ solution, let us call them $\mathbf{K3}$-type solutions,  
the value of $\epsilon$ is positive, it increases with the black hole spin $a_H$ and shows the largest variation in relative magnitude 
with an absolute total variation equal to $0.05$. 
On the other hand, for  $\mathbf{K1}$-type  and $\mathbf{K4}$-type solutions the values of epsilon are negative. 
The total variation for $\mathbf{K1}$-type solutions is equal to $0.10$ and for the $\mathbf{K4}$-type of solutions it is equal to $0.25$. 
We see that $\epsilon$ is increasing with the black hole spin $a_H$ 
for $\mathbf{K1}$-type and $\mathbf{K3}$-type solutions. It also  presents
a minimum value for $\mathbf{K4}$ at a black hole spin slightly smaller than $a_H=0$. 
This means that the magnetic collimation efficiency is lower for counter-rotating black holes in relation to their accretion disk 
for $\mathbf{K1}$-type and $\mathbf{K3}$-type solutions and it depends weakly on the spin direction
for $\mathbf{K4}$-type ones. However, $\mathbf{K4}$-type solutions are conical, contrary to all other solutions that are cylindrically collimated. 
In all cases $\epsilon$ does not vary linearly with the black hole spin $a_H$. 

This non-linear variation of epsilon with the black hole spin $a_H$ 
is explained if we try to derive it at the base of the jet where the Alfv\'enic number is equal to $0$, since some terms 
of the second order in $l$ can not be neglected in the following equation,

\begin{eqnarray}
\epsilon=  \frac{2\lambda^2}{h_z^2}\left(\frac{\Lambda^2N_B}{D} +\frac{\overline{\omega}_z}{\lambda}\right) + \lambda^2\left(\frac{\Lambda N_V}{h_* G_0 D}\right)^2 \\  \nonumber
- \frac{\nu^2(2e_1-2m_1+\delta-\kappa)R_0}{h_z^2(R_0^2+l^2)} -\frac{\nu^2l^2R_0{G_0}^2}{h_z^2(R_0^2+l^2)^3}  \\ \nonumber
\end{eqnarray}

An increasing of $\epsilon$ goes with a decreasing of the maximal Lorentz factor for the collimating Kerr solutions 
as we will see for $\mathbf{K1}$-type  and $\mathbf{K3}$-type. 
It explains why it was not possible to get physical solutions by decreasing $l$ 
for $\mathbf{K2}$-type solutions. For this second Kerr solution, we performed a fine tuning of the parameters to obtain from 
the maximally rotating black hole, the largest possible Lorentz factor at large distances.
Then decreasing $l$ leads automatically to exceed the value of the speed of light, $c$, for the polar velocity at some distance in 
the jet.
The acceleration phase does not vary for $\mathbf{K1}$-type  and $\mathbf{K3}$-type solutions with $l$ except just before $R=500$ 
where the Lorentz factor reaches a plateau. The value of the maximum $\gamma$ increases when $l$ decreases.

\begin{figure}[!h]	
\centering
\includegraphics[width=9.5cm]{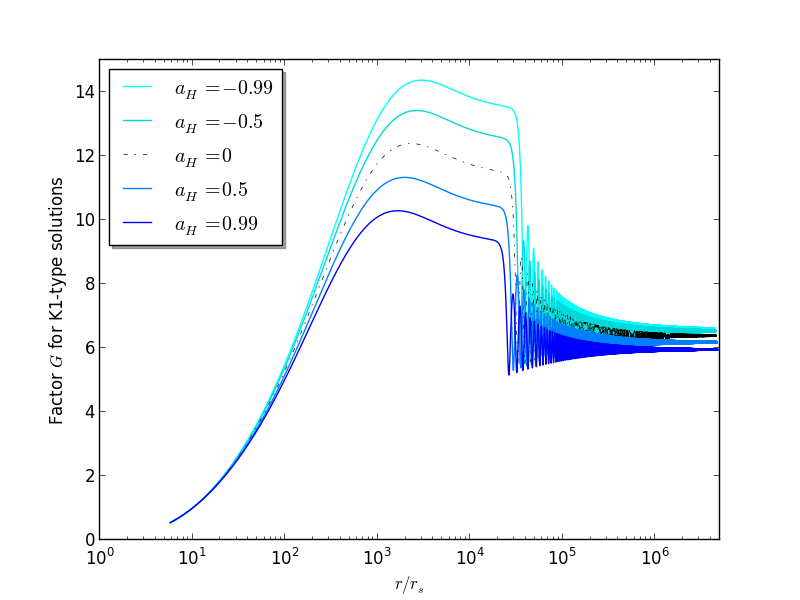}
\caption{Plot of the cylindrical jet radius normalized to its value at the Alfv\'en surface,  $G$, for $\mathbf{K1}$-type solutions, as a function of  
the distance along the polar axis, for five different values of the black hole spin $a_H$. 
The function $G$ is equal to $1$ at the Alfv\'en  distance $r=10 r_s$.}   	
\label{GCoK1}
\end{figure}

\begin{figure}[!h]	
\centering
\includegraphics[width=9.5cm]{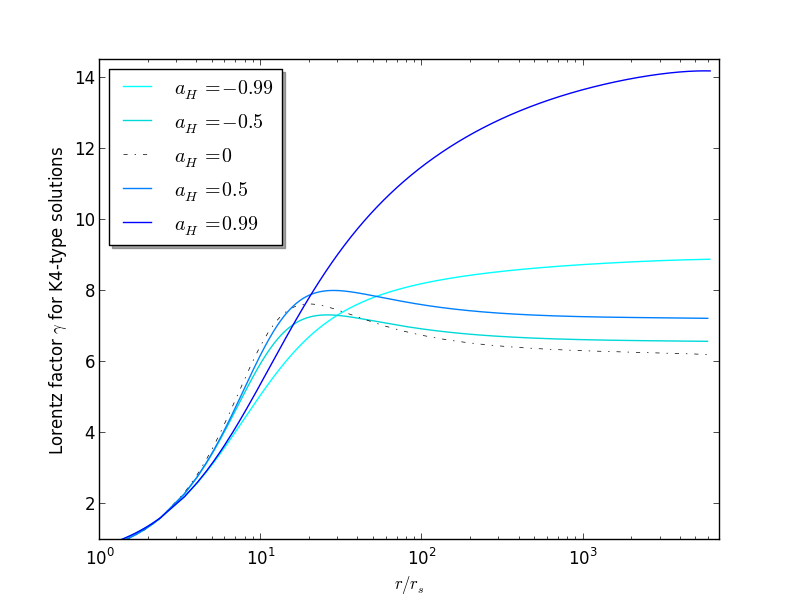}
\caption{Plot of the Lorentz factor $\gamma$ for $\mathbf{K4}$-type solutions when the black hole spin parameter $a_H$ varies between $-0.99$ and $0.99$.}   	
\label{LFK4Co}
\end{figure}

For the two collimated solutions $\mathbf{K1}$ and $\mathbf{K3}$, there is a clear effect of collimation induced by the 
rotation of the black hole. 
In Fig. \ref{GCoK1} we plot for five values of the black hole spin the evolution of the factor $G$ along the jet, i.e., the ratio of 
the jet cylindrical radius divided by its value at the Alfv\'en surface.  
From this plot it can be seen that the maximum of the radius of the jet is reached 
at different distances, as the spin varies and is decreasing when $a_H$ increases. The same trend is observed for the terminal jet radius 
but the ratio of $G_{\infty}$/$G_0$ decreases only by a factor of $0.96$ between a non-rotating and a maximally rotating black hole 
($a_H=0.99$). Hence, the faster is the black hole rotating, the smaller is the maximal jet radius. This result is expected because for 
a fast black hole rotation, the magnetic collimation efficiency parameter ($\epsilon$) is higher. 

The case of $\mathbf{K3}$-type solutions is simple, as the factor $G$ increases with the distance until it reaches a constant 
value. The ratio $G_{\infty}$/$G_0$ gives directly the expansion factor which is decreasing when $a_H$ increases from $-0.99$ up to 
$0.99$.

As it can be seen from Fig. \ref{LFK4Co}, the Lorentz factor maximum follows the opposite trend with $\epsilon$ for the conical 
$\mathbf{K4}$-type solutions. The minimum value of $\epsilon$ is obtained for $a_H \simeq 0$ and for a non-rotating black hole the Lorentz factor curve 
reaches a maximum before decreasing up to a plateau at large distances. This type of curve is not anymore observed when the 
absolute value of $a_H$ goes above  some threshold. The increase of the plateau value for the Lorentz factor is much more 
pronounced for $a_H>0$ but can be seen also for negative values of the spin. $\gamma$ at a distance of $r=1000r_s$ increases from 
a value of the order of 6 for $a_H=0$ up to $14$ for $a_H=0.99$.

\begin{figure}[!h]	
\centering
\includegraphics[width=9.5cm]{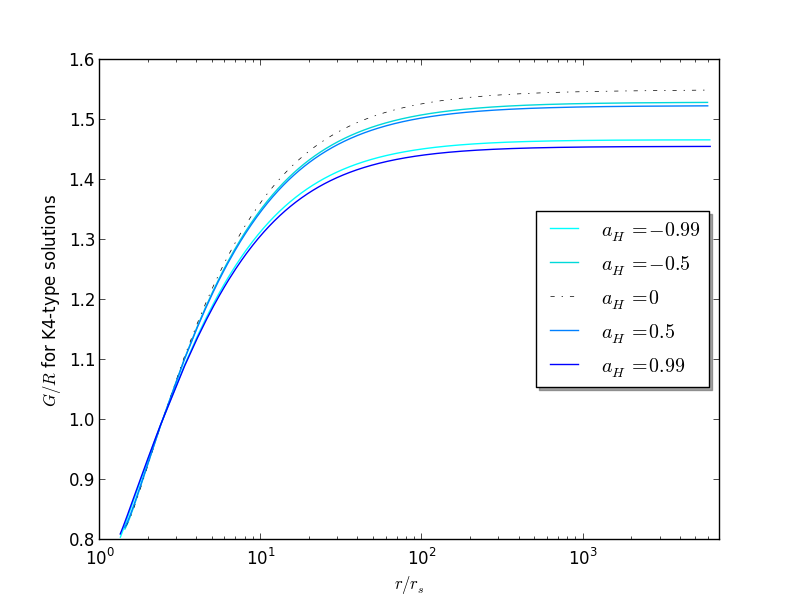}
\caption{Plot of the ratio of the cylindrical radius to the spherical radius for the conical $\mathbf{K4}$-type solutions with a spin parameter $a_H$ varying between $-0.99$ and 
$0.99$ versus the distance $z$ above the equatorial plane in units of the Schwarzschild radius. }   	
\label{ApK4Co}
\end{figure}

For the conical solutions wherein the asymptotic geometry is fixed, their geometry is less affected. The effect of varying $\epsilon$ is different in this case, as 
compared to this effect for the cylindrically collimated solutions. 
To analyze the effect of collimation induced by the black hole spin, we plot the ratio of the cylindrical to spherical radius of the field lines for  $\mathbf{K4}$-type 
solutions and five different values of the black hole spin. Note that effectively  the ratio of the cylindrical to spherical radius of the field lines gives their opening angle 
with respect to the axis. 
The field lines start at the base of the jet with an opening angle which increases with the radial 
distance and expand away from the  polar axis more rapidly for higher black hole rotation.
The effect is more pronounced for negative spin parameters. At some distances above the equatorial plane ($z \simeq 100 r_s$) 
 the opening angle becomes constant as the field lines are finally becoming radial.
It is found that for a given ratio of ${\alpha}/{\alpha_{\rm lim}}$, where $\alpha_{\rm lim}$ is the last open field line,  the asymptotic opening angle is constant for all $\mathbf{K4}$-type solutions, regardless of the spin of the black hole.

As seen in Fig. \ref{ApK4Co}, the collimation of the jet increases with $\mid a_H \mid$ which is consistent with the increase of $\epsilon$. Note also that the base of the jet slightly becomes closer to the black hole horizon as $\mid a_H \mid$  increases. 

Globally we see that the total geometry of the solution and in particular the expansion of the jet radius is extremely sensitive to the 
black hole speed. 

\subsection{The magnetic flux and power of the jets}
	
Intuitively, by physical arguments of magnetic flux conservation,  it is expected that magnetic fields play a dominant role not only in collimating large scale AGN jets, 
but also affect critically the origin of the jets in accretion disks of black hole systems, which are accordingly termed \textit{magnetically arrested disks} 
(\citealt{Nar03}).
Indeed, theoretical modeling concludes that magnetic fields at the base of AGN jets are related to the corresponding accretion rate 
(\citealt{Tchek2012}). 
\cite{Zama2014} have reported that the measured magnetic flux of the jet and the accretion disk luminosity are tightly correlated over 
several orders of magnitude for a sample of many radio-loud AGN, concluding thus that the jet launching region is threaded by a 
dynamically important magnetic field.  AGN magnetic fields can be measured either by the effect of a frequency dependent shift of the VLBI 
core position (known as the core-shift effect), or by Faraday rotation (e.g., \citealt{MartiVidaletal} who reported magnetic fields of at least tens of Gauss 
on scales of the order of several light days - 0.01 parsecs - from the black hole).  
Furthermore, magnetohydrodynamic simulations in the frame of general  relativity allow to calculate the saturation or equilibrium value for the poloidal magnetic flux $\Phi_{\rm BH}$ 
threading the black hole (\citealt{McKinney}, \citealt{Tchek2011}, \citealt{McKinney2012}). \\
In \cite{Zama2014} this poloidal flux $\Phi_{\rm BH}$ is calculated as a function of the mass accretion rate $\dot{M}$ as, 
\begin{equation}
\Phi_{\rm BH}  \simeq 50 \sqrt{\dot{M}c{\left(\frac{r_s}{2}\right)}^2}  \,.
\end{equation}
We consider that since such a strong magnetic flux can thread the black hole, we can use this formula to link our value of poloidal magnetic 
field at the Alfv\'en point $B_\star$ with the mass accretion rate $\dot{M}$. 
In our model the magnetic flux from each hemisphere is given by $\Phi_{\rm BH}=\pi \varpi_a^2 B_\star = \pi r_\star^2 B_\star \alpha_{\rm lim}$. 
Then, the magnitude of the magnetic field is calculated from the expression, 
\begin{equation}
B_\star \simeq 25 r_s \frac{\sqrt{\dot{M}c}}{\pi r_\star^2 \alpha_{\rm lim}} = 25 \mu \frac{\dot{M}^{1/2}c^{1/2}}{\pi r_\star \alpha_{\rm lim}} \,.
\end{equation}

In order to compare the jet power for the \textbf{K2} and \textbf{K3} solutions with the one obtained by general relativistic magnetohydrodynamic  simulations, we calculate this power in terms of the parameters of our model. 
Similarly to the way we deduced the angular momentum flux density in Eqs. (\ref{AMF1}) and (\ref{AMF2}), we may calculate the 
jet+counterjet power,  by substituting  $\Psi_A$ from Eq. (\ref{defPsiA}), ${\cal E} = {\cal E} _\star(1+e_1 \alpha)$ with 
$\cal{E_\star}$  from Eq. (\ref{e_E}) and with the help of Eq. (\ref{Vstar}), we obtain in terms of the constants of our model,  
\begin{eqnarray}\nonumber
P_{\rm jet}&=& \int_0^{A_{\rm lim}} \Psi_A {\cal E} dA \\   
&=& \frac{\nu h_\star^2 c}{2 \sqrt{\mu}} (B_\star r_\star)^2 \int_0^{\alpha_{\rm lim}} (1+e_1 \alpha) \sqrt{1+\delta \alpha} d\alpha\,  
\end{eqnarray}
Hence, we finally get,
\begin{equation}
 P_{\rm jet}\simeq 625 \frac{\nu \mu^{3/2} h_\star^2}{2 \pi^2 \alpha_{\rm lim}^2}  \dot{M} c^2     \int_0^{\alpha_{\rm lim}} (1+e_1 \alpha) \sqrt{1+\delta \alpha} d\alpha \,
\end{equation}

Therefore, we get for the efficiency $\eta_{\rm jet} \equiv {P_{\rm jet}}/{\dot{M}c^2}$  for our \textbf{K2} model a value  $\eta_{\rm jet} \simeq 0.52$, while for our 
\textbf{K3} model $\eta_{\rm jet} \simeq 0.40$. 
On the other hand, \cite{McKinney} determined self-consistently the jet power in Blandford-Znajek numerical models and 
deduced an efficiency $\eta_{\rm jet}$ between 0.01 and 0.1 for ultra-relativistic 
Poynting-dominated jets with $a_H$ larger than 0.8. 
Later on,  \cite{Tchek2011} and \cite{McKinney2012} increased the magnetic flux which can be pushed near the black hole leading 
to magnetically arrested accretion and obtained values of the net flow efficiency larger than 1 for rapidly spinning black holes 
with $a_H$ larger than 0.9. 
Their models that develop a highly non-axisymmetric magnetically choked accretion flow, have initially the poloidal component of the magnetic field
 dominant and the wind has an efficiency always smaller than the one of the jet. 
Note that the \textbf{net} flow efficiency for the jet is equal to $ (P_{\rm jet}-\dot{M}_Hc^2) /  [\dot{M}_Hc^2]_t$ where $\dot{M}_H$ is the 
black hole mass accretion rate and $[\dot{M}_Hc^2]_t$ is the time averaged value of accretion power. This black hole mass accretion rate 
could be smaller than the mass accretion rate measured by \cite{Zama2014}.  In fact, they deduced the accretion power by dividing the bolometric luminosity with  
a radiative efficiency of 0.4. Larger values of the inflow rates $\dot{M}_{\rm in, i}$ and $\dot{M}_{\rm in, o}$ have been obtained 
by \cite{McKinney2012} at radii $5r_s$ and $25r_s$, respectively.
\cite{Zdziarski2015} found also that the jet power exceeds moderately the accretion power $\dot{M}c^2$ for blazars estimating the 
magnetic flux from the radio-jet core shift effect and the self-absorbed flux evaluation. 
However, there is a large scatter around the mean value for blazars $\eta_{\rm jet}  \simeq 1.3$ 
and the jet power for radio galaxies 
is smaller, especially for M87. Then our estimations of jet power from our \textbf{K2} and \textbf{K3} solutions with mass loading could 
perfectly match that for less efficient blazars and radio galaxies.
Moreover, they are not depending too much on the spin parameter $a_H$, since ${\alpha_{\rm lim}}$ keeps a value slightly smaller than 1 
when the spin parameter varies for the  \textbf{K3}-type solutions, even for retrograde black holes. \\ 
At this point, we prefer to postpone a further discussion of the jet power, until we shall have completed our study, which also includes inflow solutions and leads to a 
spin-energy extraction or addition from the black hole (work in progress).

\section{Summary and Conclusions}

As it was pointed already in 1957 by Parker (see also  \citealt{Parker63}), for the driving of the solar wind and similar enthalpy driven astrophysical outflows 
some energy/momentum addition is required. 
The original isothermal and polytropic models with a heat conduction, have shown that effectively energy and/or momentum are necessary for producing 
supersonic/superAlfvenic outflows at large distances, to also meet the respective causality requirements.    
Quasi-radial wind-type astrophysical outflows with shock transitions (\citealt{HabbalTsinganos83}) have been applied to explain the appearance of emission knots 
in galactic (\citealt{Silvestroetal87}) and extragalactic objects (\citealt{Ferrarietal86}), also in the framework of special relativity (\citealt{Ferrarietal85}). \\
When deviations of the outflow geometry from radial expansion exist and the problem is fully 2-dimensional, with a suitable external gas pressure distribution and mainly by 
magnetic fields, these outflows can be collimated in the form of jets (\citealt{ST94}).   
Along these lines, in \cite{VlahakisTsinganos98} the original Parker model has been extended to include general MHD effects, in the context of meridional 
self-similarity. 
The present paper takes the extra step of using the framework of a Kerr metric to explore analogous enthalpy- or generalized pressure-driven outflows from 
the environment of a rotating black hole. 

Specifically, in this paper we presented an exact MHD solution for an outflow in a Kerr metric, constructed by using the assumption of self similarity and the 
mechanism for driving the outflow which is developed in \cite{ST94}. 
Additionally, the model is based on a first order expansion of the governing general relativistic equations in the magnetic flux function around the symmetry axis of the system. It yields four nonlinear and coupled differential equations as a function of the radius, for the Alfv\'en number, the gas pressure, the expansion function and the 
radius of the jet.
 The model depends on seven parameters. Two of them are the meridional increase of the gas pressure and the mass to magnetic flux ratio, $\kappa$ and 
$\delta$ respectively. There is also the meridional increase of the total energy with the magnetic flux function, $e_1$, the poloidal current density flowing along the 
system axis, $\lambda$, the escape speed in units of the Alfv\'en speed, $\nu$,  the Schwarzschild radius in units of the Alfv\'en radius, $\mu$ and the 
dimensionless black hole spin $l$.  
In addition to those seven parameters, we have to adjust the pressure at the Alfv\'enic transition. We chose to adjust it such as to 
minimize the  oscillations of the magnitude of the flow speed along the axis and taking the limiting solution. We also fix 
the magnetic field at the Alfv\'en transition,  $B_\star$, and a uniform pressure constant $P_0$ to ensure a zero external heating at infinity along the axis where 
the Lorentz factor is maximum. 

The model takes into account the light cylinder effects and the meridional increase of the Alfv\'en number with the magnetic flux function, $m_1$. This parameter is deduced from the regularity conditions at the Alfv\'en transition surface.

The classical energetic criterion for the transition from conical winds to cylindrical jets is generalized in general relativity and it amounts to say that 
if the total {\textit{available}} energy along a nonpolar streamline exceeds the corresponding energy along the axis, then the outflow collimates in a jet. 

In the framework of a Kerr metric, we illustrate the model with four different enthalpy driven solutions wherein the contribution of the Poynting flux is rather small. 
The first three solutions are cylindrically collimated, while the fourth represents a conical outflow at infinity. 
The flow collimation is induced by electromagnetic forces. In all four models, relativistic speeds are obtained while in one of them the Lorentz factor 
$\gamma$ obtains ultra relativistic values.
A preliminary application of one of our Kerr solutions ($\mathbf{K3}$) was explored to model the spine jet in M87, yielding encouraging results. 
A more complete modeling for the M87 jet including an external disk-wind component  will be explored in another connection.  \\ 

Our analytical solutions of the full general relativistic MHD equations in a Kerr metric may contribute to a better understanding of relativistic AGN jets 
and are complementary to sophisticated numerical simulations of such jets (e.g., \citealt{McKinney}, \citealt{Tchek2011}, \citealt{McKinney2012}).  
In both approaches, the analytical and the numerical one,  the outflows are electromagnetically confined. 
However, while in the above numerical simulations the outflow is driven electromagnetically (e.g., via the Blandford-Znajek mechanism), 
in the present analytical solutions the outflow from the hot corona surrounding the black hole is enthalpy- or generalized pressure-driven  
(e.g, via the Sauty-Tsinganos mechanism).    
Nevertheless, it is interesting to note that the jet power for the two representative analytical solutions we presented in this paper is similar to the ones   
determined by the numerical simulations.
 
The present model can also serve to construct an inflow solution, in order to link it with an outflow solution and the physical creation of leptonic pairs 
to determine the energy balance of the black hole, via a generalized Penrose process as compared to the Blandford-Znajek mechanism. This undertaking 
is in progress and it will  be presented in another connection.

\begin{acknowledgements}
      Part of this work was supported by the French
      \emph{Action F\'ed\'eratrice CTA} project and the "Programme Blanc" of Paris Observatory. 
      The authors thank the referee for his constructive suggestions which helped a better presentation 
      of the results of the paper, in particular for a comparison with the results of numerical simulations. 
\end{acknowledgements}

\appendix
\onecolumn

\section{Kerr metric elements}
\label{appendixA}

We can rewrite the Kerr-metric in a simpler form,
\begin{equation}
ds^2 = -h^2c^2{dt}^2 + {h_r}^2{dr}^2 + {h_{\theta}}^2{d\theta}^2+{h_{\phi}}^2\left(d\phi+\beta^{\phi}cdt\right)^2
\,.
\end{equation}
The Kerr metric elements are defined as,
\begin{eqnarray}
h_r=\dfrac{\rho}{\sqrt{\Delta}}\,,
&h_{\theta}=\rho \,,
\\
h_{\phi}=\varpi=\dfrac{\Sigma}{\rho}\sin \theta\,,
&\omega=\dfrac{c a r_s r}{\Sigma^2}\,,
\\
\beta_{\phi}=-\dfrac{\omega}{c}\varpi^2\,,
&\beta^{\phi}=-\dfrac{\omega}{c}\,, \\
h=\left(1-\dfrac{r_s r}{\rho^2}+\beta^{\phi}\beta_{\phi}\right)^{1/2}
&=\dfrac{\rho}{\Sigma}\sqrt{\Delta}
%\,, &\vec{\beta}=\beta^{\phi} \vec{e}_{\phi}
\,.
\end{eqnarray}
%
%where $(\vec{e}_i)_{i=1...4}$ is the natural basis. 

Thus, to second order in $\sin \theta$ the metric is,		
 \begin{eqnarray}
\omega & = & \frac{lc\mu R}{r_\star(R^2+l^2)^2} \left(1+\frac{l^2 {h_z}^2}{R^2+l^2}\sin^2\theta \right) \\
h_{\,} & = & \sqrt{1-\frac{\mu R}{R^2+l^2}}\left(1-\frac{\mu l^2 R}{2(R^2+l^2)^2}\sin^2\theta \right)\\
h_r&=&\frac{1}{\displaystyle \sqrt{1-\frac{\mu R}{R^2+l^2}}}\left(1-\frac{l^2}{2(R^2+l^2)}\sin^2\theta \right)\\
h_{\theta} & = & r_\star \sqrt{R^2+l^2}\left(1-\frac{l^2 }{2(R^2+l^2)}\sin^2\theta \right)\\
h_{\phi} & = & \varpi =  r_\star\sqrt{R^2+l^2}\sin\theta 
\,.
\label{OHDEF}
\end{eqnarray}    

In order to simplify our notation, we define along the polar axis, the axial lapse function,
 \begin{equation}
 h_z(R) = h_r^{-1}(R,\theta=0) = h(R,\theta=0)= \sqrt{1-\frac{\mu R}{R^2+l^2}}
\label{hzdef}
\end{equation}    	
 and the axial shift of the metric, 
 \begin{equation}
\omega_z(R)=\omega(R,\theta=0)=\frac{lc\mu R}{r_\star(R^2+l^2)^2}
\,.
\label{omzdef}
\end{equation}  	  

%%%%%% JUST ADD IT YOU CAN CUT IT IF NECESSARY
We can also write the metric as an expansion in $\alpha$. 
 \begin{eqnarray}
\omega & = & \frac{lc\mu R}{r_\star(R^2+l^2)^2} \left(1+\frac{l^2 {h_z}^2 G^2}{(R^2+l^2)^2}\alpha \right) \\
h_{\,} & = & \sqrt{1-\frac{\mu R}{R^2+l^2}}\left(1-\frac{\mu l^2 R G^2}{2(R^2+l^2)^3}\alpha \right)\\
h_r&=&\frac{1}{\displaystyle \sqrt{1-\frac{\mu R}{R^2+l^2}}}\left(1-\frac{l^2 G^2}{2(R^2+l^2)^2}\alpha \right)\\
h_{\theta} & = & r_\star \sqrt{R^2+l^2}\left(1-\frac{l^2 G^2}{2(R^2+l^2)^2}\alpha \right)\\
h_{\phi} & = & \varpi =  r_\star G \sqrt{\alpha}
\,.
\label{OHDEF}
\end{eqnarray}    
%%%%%%%%%%

\vspace{1cm}

\section{Vectorial Operators in Boyer-Lindquist coordinates of the Kerr metric.}
\label{appendixB}

In this Appendix we summarize the expressions of the vectorial operators under the assumption of axisymmetry ($\partial_\phi=0$).
\\
\par\noindent
1. The gradient vector $\vec{\nabla}$ in the ZAMO orthonormal bases,

\begin{equation}
\vec{\nabla} = \sum_{i=1}^{3}\frac{\vec{\epsilon}_i}{h_i}\partial_i 
\label{gradexp}
\end{equation}   

\par\noindent
2. The divergence of a vector $\vec{V}$,
    
\begin{equation}
\vec{\nabla}\cdot\vec{V}=\frac{1}{h_r h_{\theta} \varpi} \left[\partial_r(h_{\theta} \varpi V^{\hat{r}})+\partial_{\theta}(h_r \varpi V^{\hat{\theta}})\right]
\label{Divexpr}
\end{equation}

\par\noindent
3. The scalar Laplace operator,
            
\begin{eqnarray*}
\nabla^2 A&=& \vec{\nabla}\cdot\vec{\nabla A}\\
&=&\frac{1}{h_r h_\theta \varpi}\left[\frac{1}{r_\star^2}\partial_R\left(\frac{h_\theta \varpi}{h_r}\partial_R A\right)+\partial_\theta\left(\frac{h_r \varpi}{h_\theta}\partial_\theta A\right)\right]
\end{eqnarray*}

\par\noindent       
4. The curvature operator on a vector $\vec{V}$,
    
\begin{eqnarray*}
\vec{\nabla} \times \vec{V}  = % ^3\epsilon^i_j^k\partial_i(V^j)\vec{e}_k\\
  \frac{h_k}{h_r h_{\theta} h_{\phi}} \epsilon^{ijk}\partial_i(h_j V^{\hat{j}})\vec{\epsilon}_k
\end{eqnarray*}

%Thus,
\begin{equation}
\vec{\nabla} \times \vec{V} =
 \begin{bmatrix}
 \frac{1}{h_{\theta}\varpi}\partial_{\theta}(\varpi {V}^{\hat{\phi}}) \\
- \frac{1}{h_r \varpi r_*}\partial_R (\varpi V^{\hat{\phi}}) \\
 \frac{1}{h_r h_{\theta}}\left(\frac{1}{r_*}\partial_R(h_{\theta} {V}^{\hat{\theta}})-\partial_{\theta}(h_r V^{\hat{r}}) \right)
 \end{bmatrix}
 \label{rotational}
\end{equation}
\\
\par\noindent    
5. The advection term,
    
\begin{equation*}
(\vec{V}\cdot\vec{\nabla})\vec{V}=V^{\alpha}D_{\alpha}V^{\beta} \vec{e}_{\beta}
\end{equation*}
After some algebra we get for the poloidal component of the advection term:
\begin{equation}
[(\vec{V}\cdot\vec{\nabla})\vec{V}]_{\rm p}=
 \begin{bmatrix}
 \frac{V^{\hat{r}}}{r_*h_r}\partial_R V^{\hat{r}}+\frac{{V}^{\hat{\theta}}}{h_{\theta}}\partial_{\theta} {V}^{\hat{r}}+\frac{V^{\hat{r}}V^{\hat{\theta}}}{h_{\theta}}\partial_{\theta} ln(h_r)-\frac{(V^{\hat{\theta}})^2}{h_r}\partial_r ln(h_{\theta})-\frac{(V^{\hat{\phi}})^2}{h_r}\partial_r ln(h_{\phi})\\
 \frac{V^{\hat{r}}}{r_*h_r}\partial_R V^{\hat{\theta}}+\frac{V^{\hat{\theta}}}{h_{\theta}}\partial_{\theta} V^{\hat{\theta}}+\frac{V^{\hat{r}}V^{\hat{\theta}}}{r_* h_r}\partial_R ln(h_{\theta})-\frac{(V^{\hat{r}})^2}{h_{\theta}}\partial_{\theta} ln(h_r)-\frac{(V^{\hat{\phi}})^2}{h_{\theta}}\partial_{\theta} ln(h_{\phi})
 \label{Advec}
 \end{bmatrix}
\end{equation}
It can be useful to get the non-symmetric advection term,
\begin{equation}
[(\vec{B}\cdot\vec{\nabla})\vec{C}]=
 \begin{bmatrix}
 \frac{{B}^{\hat{r}}}{r_*h_r}\partial_R C^{\hat{r}}+\frac{B^{\hat{\theta}}}{h_{\theta}}\partial_{\theta} C^{\hat{r}}+\frac{B^{\hat{r}} C^{\hat{\theta}}}{h_{\theta}}\partial_{\theta} ln(h_r)-\frac{B^{\hat{\theta}}C^{\hat{\theta}}}{h_r}\partial_r ln(h_{\theta})-\frac{B^{\hat{\phi}}C^{\hat{\phi}}}{h_r}\partial_r ln(h_{\phi})\\
 \frac{B^{\hat{r}}}{r_*h_r}\partial_R C^{\hat{\theta}}+\frac{B^{\hat{\theta}}}{h_{\theta}}\partial_{\theta} C^{\hat{\theta}}+\frac{C^{\hat{r}}B^{\hat{\theta}}}{r_* h_r}\partial_R ln(h_{\theta})-\frac{B^{\hat{r}}C^{\hat{r}}}{h_{\theta}}\partial_{\theta} ln(h_r)-\frac{B^{\hat{\phi}}C^{\hat{\phi}}}{h_{\theta}}\partial_{\theta} ln(h_{\phi})\\
 \frac{B^{\hat{r}}}{h_r r_\star}\partial_R C^{\hat{\phi}}+\frac{B^{\hat{\theta}}}{h_\theta}\partial_\theta C^{\hat{\phi}}+\frac{C^{\hat{r}}B^{\hat{\phi}}}{r_* h_r}\partial_R ln(h_{\phi})+\frac{C^{\hat{\theta}}B^{\hat{\phi}}}{h_\theta}\partial_R ln(h_{\phi})
 \label{Advec}
 \end{bmatrix}
\end{equation}

\section{Final differential equations of the MHD problem}
\label{appendixC}

The final ordinary differential equations of our model can be written as :
 \begin{equation}
% \fbox{$
   \begin{array}{rcl}
\displaystyle{R\frac{d}{dR}}
\left( \begin{array}{c}
M^2 \\
F 
%\Pi 
\end{array} \right)=
\frac{1}{\mathcal{D}(M^2,G^2,F,R)} 
\left( \begin{array}{c}
\mathcal{N}_{M^2}  \\
\mathcal{N}_{F}  \\
%\mathcal{N}_{\Pi} 
\end{array} \right)
   \end{array}   \,,%$}
   \label{FS}
\end{equation}
% \begin{equation}
%\fbox{$
%   \begin{array}{rcl}
%\displaystyle{R\frac{d}{dR}}
%\left( \begin{array}{c}
%M^2 \\
%F \\
%\Pi 
%\end{array} \right)=
%\frac{1}{\mathcal{D}(M^2,G^2,F,\Pi,R)} 
%\left( \begin{array}{c}
%\mathcal{N}_{M^2}  \\
%\mathcal{N}_{F}  \\
%\mathcal{N}_{\Pi} 
%\end{array} \right)
%   \end{array}   \,,$}
%   \label{FS}
%\end{equation}
where
\begin{equation}
%   \fbox{$
   \begin{array}{rcl}
\mathcal{D}(m^2,G^2,F,R)
&=&\displaystyle{
\frac{h_*^2}{R}} \left[-D\left(1+(\kappa-2 m_1)\frac{R^2+l^2}{G^2}-\frac{l^2}{R^2 +
l^2} \right) \right.+
%&+&  \left.  
%\displaystyle{ 
\left. \frac{\lambda^2\Lambda^2 N_B^2 (R^2+l^2)}{D^2}+\frac{h_z^4F^2(R^2+l^2)}{4
h_*^2 R^2} \right] 
   \end{array}  \,, %$}
   \label{SubAlf}
\end{equation}
and,
 \begin{equation}
%   \fbox{$
   \begin{array}{rcl}
\mathcal{N}_{F}&=& \displaystyle{ 
    \frac{FM^2}{h_*^2}\left[\frac{F}{2}\left(\frac{h_z^2F}{2}-1\right)+\left(\frac{F}{2}-1\right)\left(1+(\kappa-2m_1)\frac{X_{+}}{G^2}-\frac{l^2}{X_{+}}+\frac{\lambda^2\Lambda^2 N_B^2X_{+}}{D^3} \right) \right] } \\
&&\\
&+& \displaystyle{
\frac{R^2 h_z^2}{X_{+}h_*^2}\left[\frac{X_{+}}{R^2}F(F-1)-\frac{2}{h_z^2}-\frac{4\lambda^2\mu h_*^2\Lambda^2X_{+}}{\nu^2 h_z^4}-\frac{4l^2\mu R}{h_z^2X_{+}^2}\right]\left[1+(\kappa-2m_1)\frac{X_{+}}{G^2}-\frac{l^2}{X_{+}}-\frac{\lambda^2\Lambda^2N_B^2X_{+}}{D^3}-\frac{h_z^2F^2X_{+}}{4R^2}\right]  } \\ 
&&\\
&+& \displaystyle{
\left(\frac{2\kappa \Pi G^2 R^2}{h_*^2}+\frac{\mu F R X_{-}}{h_*^2 X_{+}^2}\right) \bigg{[}1+(\kappa-2m_1)\frac{X_{+}}{G^2}-\frac{l^2}{X_{+}}-\frac{\lambda^2\Lambda^2N_B^2X_{+}}{D^3}-\frac{h_z^2F^2X_{+}}{4R^2}\bigg{]}  }\\
&&\\
&+&\displaystyle{
\frac{\nu^2 h_*^2 F G^2 R}{2 h_z^2 M^2}\frac{X_{-}}{X_{+}}(\kappa-2e_1+2m_1-\delta)-\frac{\mu F M^2 R X_{-}}{2 h_*^2 h_z^2 X_{+}^2}\left[1+(\kappa-2m_1)\frac{X_{+}}{G^2}-\frac{l^2}{X_{+}}\right] }\\
&&\\
&+& \displaystyle{
\frac{\lambda^2\mu\Lambda^2 N_B N_V F R}{h_*^2 D^3}\frac{X_{-}}{X_{+}}-\frac{\lambda^2 \Lambda N_B h_z^2 F X_{+}}{h_*^2 D^2}\left(F-\frac{2R^2}{X_{+}}\right)
+\frac{\lambda^2 \mu h_*^2 F G^2}{h_z^2 M^2}\frac{RX_{-}}{X_{+}}\left(\frac{\Lambda^2 N_B}{D}+\frac{\bar{\omega}_z}{\lambda}\right) }\\
&&\\
&+& \displaystyle{
\frac{4 \lambda^2 \Lambda^2 R^2}{h_z^2}\left(\frac{N_B^2}{D^2}-\frac{h_z^2}{2M^2}\frac{N_V^2}{D^2}\right)\left[1+(\kappa-2m_1)\frac{X_{+}}{G^2}-\frac{l^2}{X_{+}}-\frac{\lambda^2 \Lambda^2 N_B^2X_{+}}{D^3}-\frac{h_z^2F}{2}\right] }\\
&&\\
&+& \displaystyle{
\lambda \sqrt{\mu}\nu l h_*\frac{\Lambda G^2 F R}{D}\frac{(3R^2-l^2)}{X_{+}^2}\left(\frac{N_B}{h_*^2 D}-\frac{N_V}{M^2}\right) }\\
&&\\
&-& \displaystyle{
\frac{2 \nu^2 l^2 h_*^2G^4}{h_z^2 M^2}\frac{R^3}{X_{+}^3}\left[1-\frac{l^2}{X_{+}}+(\kappa-2m_1)\frac{X_{+}}{G^2}-\frac{\lambda^2 \Lambda^2 N_B^2 X_{+}}{D^3}+\left(\frac{X_{-}}{4R^2}+\frac{h_z^2}{2}\right)F\right] }\\
&&\\
&+& \displaystyle{
\frac{2M^2 l^2 R^2}{h_*^2 h_z^2 X_{+}^2}\left[1+(\kappa-2m_1)\frac{X_{+}}{G^2}-\frac{l^2}{X_{+}}-\frac{\lambda^2\Lambda^2N_B^2X_{+}}{D^3}+\frac{h_z^2 FX_{+}}{R^2}\left(\frac{3R^2}{2X_{+}}+(\kappa-2m_1)\frac{X_{+}}{G^2}\right)\right] }
   \end{array}   %$}
   \label{NF}
   \end{equation}
   where 
 $ X_+ = R^2 + \ell^2\;\;  X_- = R^2 -\ell^2$.

\begin{equation}
%   \fbox{$
   \begin{array}{rcl}
\mathcal{N}_{M^2}
&=& \displaystyle{
\frac{M^4}{4 h_*^2}\left[-h_z^2 F^2+2F-\frac{4R^2l^2}{X_{+}^2}-4\left(F-2\frac{R^2}{X_{+}}\right)\left(1+(\kappa-2m_1)\frac{X_{+}}{G^2}-\frac{l^2}{X_{+}}\right)\right]}\\
&&\\
&+&  \displaystyle{
\frac{h_z^2M^2}{h_*^2}\left[\frac{h_z^2X_{+}F^3}{8 R^2}+\frac{h_z^2F^2}{4}\left(1+\frac{\mu X_{-}}{h_z^2RX_{+}}\right)+(\kappa-2m_1)\frac{X_{+}F}{G^2}-F\frac{\lambda^2\mu}{\nu^2}X_{+}\Lambda^2\frac{h_*^2}{h_z^2}-2\frac{R^2}{X_{+}}-(\kappa-2m_1)\frac{2R^2}{G^2}+\frac{3R^2l^2}{X_{+}^2}-\frac{Fl^2}{X_{+}}\left(\frac{3}{2}-h_z^2\right)\right]}\\
&&\\
&+& \displaystyle{
\frac{\nu^2 h_*^4 D RG^2}{2h_z^2M^2}\frac{X_{-}}{X_{+}}\left(\kappa-\delta+2m_1-2e_1\right)+\kappa \frac{X_{+}}{2}\frac{h_z^2}{h_*^2} F \Pi G^2 M^2-\frac{DM^2\mu RX_{-}}{2h_z^2X_{+}^2}\left[1+(\kappa-2m_1)\frac{X_{+}}{G^2}-\frac{l^2}{X_{+}}\right]+\frac{\lambda^2 \mu R\Lambda^2 N_B N_V}{D^2}\frac{X_{-}}{X_{+}}}\\
&&\\
&+& \displaystyle{
\lambda^2 \Lambda^2 X_{+}\left(\frac{N_B^2}{D^2}-\frac{h_z^2}{2M^2}\frac{N_V^2}{D^2}\right)\left(2M^2\frac{R^2}{X_{+}}+h_z^2(F-2\frac{R^2}{X_{+}})\right)-\frac{\lambda^2\Lambda N_B h_z^2X_{+}}{D}\left(F-2\frac{R^2}{X_{+}}\right)+\frac{\lambda^2\mu h_*^4RG^2}{h_z^2M^2}\frac{X_{-}}{X_{+}}\left(\Lambda^2N_B+\frac{Dr_* \omega_z }{\lambda V_* h_*}\right) }\\
&&\\
&-& \displaystyle{
\frac{l^2\nu^2 R G^4h_*^2}{2M^2 X_{+}^3}\left[h_*^2 D\left(3R^2-l^2+\frac{\mu R X_{-}}{h_z^2 X_{+}}\right)+h_z^2FX_{+}\right]+{\lambda \sqrt{\mu}\nu l h_* RG^2 \Lambda} \frac{l^2-3R^2}{X_{+}^2}\left(\frac{N_V h_*^2}{M^2}-\frac{N_B}{D}\right) }
   \end{array}
%   $}
   \label{NM}
\end{equation}

\begin{equation}
   \begin{array}{rcl}
\displaystyle{\frac{d \Pi}{dR}}=
\displaystyle{ -\frac{2}{h_z^2G^4}\left[\frac{d}{dR}M^2+\frac{M^2}{R}\left(F-\frac{2R^2}{R^2+l^2}\right)\right]-\frac{1}{h_z^4M^2}\frac{R^2-l^2}{(R^2+l^2)^2}\left(\nu^2h_*^4-\frac{\mu M^4}{G^4}\right)}
   \end{array}   \,,
\label{Pressure}
\end{equation}

\begin{equation}
   %\fbox{$
   \begin{array}{rcl}
\displaystyle{\frac{d G^2}{dR}}=\displaystyle{\frac{G^2}{R}\left(\frac{2R^2}{R^2+l^2}-F\right)}
   \end{array}   \,.%$}
\label{G-F}
\end{equation}


\vspace{1cm}

\begin{thebibliography}{}

\bibitem[\protect\citeauthoryear{Anderson et al.}{2003}]{Andersonetal03} 
Anderson J.~M., Li Z.-Y., Krasnopolsky R., Blandford R.~D., 2003, ApJ, 590, L107 

\bibitem[\protect\citeauthoryear{Begelman et al.}{2008}]{Begelman08} 
Begelman M.~C., Fabian A.~C., Rees M.~J., 2008, \mnras, 384, L19

\bibitem[\protect\citeauthoryear{Blandford \& Payne}{1982}]{BlandfordPayne82}
Blandford, R. D., \& Payne, D. G. 1982,  \mnras, 199, 883

\bibitem[\protect\citeauthoryear{Blandford \& Znajek}{1977}]{BlandfordZnajek77}
Blandford, R. D., \& Znajek, R. L. 1977,  \mnras, 179, 433

\bibitem[\protect\citeauthoryear{Beskin}{2010}]{beskin10} 
 Beskin V.~S. 2010, 
     MHD Flows in Compact Astrophysical Objects: Accretion,
     Winds and Jets (Springer-Verlag Berlin Heidelberg)

\bibitem[\protect\citeauthoryear{B{\"o}ttcher et al.}{2013}]{Boettcheretal13}
B{\"o}ttcher, M., Reimer, A., Sweeney, K. and Prakash, A. 2013, ApJ, 768, 54

\bibitem[\protect\citeauthoryear{Breitmoser  \& Camenzind}{2000}]{Breitmoser2000} 
Breitmoser, E. \& Camenzind, M.\ 2000, \aap, 361, 207 

\bibitem[\protect\citeauthoryear{Casadio et al.}{2015}]{Casadio15} 
Casadio C., et al., 2015, ApJ, 813, 51

\bibitem[\protect\citeauthoryear{Cayatte et al.}{2014}]{cayatteetal14}
Cayatte, V., Vlahakis, N., Matsakos, T., Lima, J. J. G., Tsinganos, K. \& Sauty, C. 
2014, ApJL, 788, L19

\bibitem[\protect\citeauthoryear{Celotti \& Ghisellini}{2008}]{CelottiGhisellini08}
Celotti, A. \& Ghisellini, G. 2008, \mnras, 385, 283

\bibitem[\protect\citeauthoryear{Fabian \& Rees}{1995}]{FabianRees95}
Fabian, A.C., \& Rees, M.J. 1995, \mnras, 277, L55

\bibitem[\protect\citeauthoryear{Fenget al.}{2016}]{Feng16} 
Feng J., Wu Q., Lu R.-S., 2016, ApJ, 830, 6

\bibitem[\protect\citeauthoryear{Ferrari et al.}{1985}]{Ferrarietal85}
Ferrari, A., Trussoni, E., Rosner, R. \& Tsinganos, K. 1985, ApJ, 294, 397

\bibitem[\protect\citeauthoryear{Ferrari et al.}{1986}]{Ferrarietal86}
Ferrari, A., Trussoni, E., Rosner, R. \& Tsinganos, K. 1986, ApJ, 300, 577

\bibitem[\protect\citeauthoryear{Finke}{2016}]{Finke16}
Finke, J. D. 2016, ApJ, 830, 94

\bibitem[\protect\citeauthoryear{Giroletti et al.}{2004}]{Giroletti04} 
Giroletti M., et al., 2004, ApJ, 600, 127

\bibitem[\protect\citeauthoryear{Ghisellini et al.}{2005}]{Ghisellini05} 
Ghisellini G., Tavecchio F., Chiaberge M., 2005, A\&A, 432, 401

\bibitem[\protect\citeauthoryear{Globus et al.}{2014}]{Globusetal14} 
Globus, N., Sauty C., Cayatte V.  \& Celnikier, L.-M. 2014,
      PhRvD, 89l4015

\bibitem[\protect\citeauthoryear{Gracia et al.}{2009}]{Gracia09}
Gracia J., Vlahakis N., Agudo I., Tsinganos K., Bogovalov S.~V., 2009, ApJ, 695, 503

\bibitem[\protect\citeauthoryear{Hada et al.}{2011}]{Hada11} 
Hada K., Doi A., Kino M., Nagai H., Hagiwara Y., Kawaguchi N., 2011, Natur, 477, 185

\bibitem[\protect\citeauthoryear{Habal \& Tsinganos}{1983}]{HabbalTsinganos83} Habal, S.R., \& Tsinganos, K., JGR, 88, 1965 

\bibitem[\protect\citeauthoryear{Hervet et al.}{2017}]{Hervetetal17}
Hervet, O., Meliani, Z., Zech, A., Boisson, C., Cayatte, V., Sauty, C. and Sol, H. 2017, A\&A, 

\bibitem[\protect\citeauthoryear{Komissarov}{2007}]{Komissarov07} 
Komissarov S.~S., 2007, MNRAS, 382, 995

\bibitem[\protect\citeauthoryear{Kravchenko et al.}{2016}]{Kravchenko16} 
Kravchenko E.~V., Kovalev Y.~Y., Hovatta T., Ramakrishnan V., 2016, MNRAS, 462, 2747

\bibitem[\protect\citeauthoryear{Li et al.}{2009}]{Lietal09} Li Y.-R., Yuan Y.-F., Wang J.-M., Wang J.-C., Zhang S., 2009, ApJ, 699, 513 

\bibitem[\protect\citeauthoryear{Marscher et al.}{2010}]{Marscher10}
Marscher, A. P. et al., 2010, ApJL, 710, 126

\bibitem[\protect\citeauthoryear{Mart{\'{\i}} et al.}{2016}]{Marti16} 
Mart{\'{\i}} J.~M., Perucho M., G{\'o}mez J.~L., 2016, ApJ, 831, 163

\bibitem[\protect\citeauthoryear{Mart{\'{\i}}-Vidal et al.}{2015}]{MartiVidaletal} Mart{\'{\i}}-Vidal I., Muller S., Vlemmings W., Horellou C., Aalto S., 2015, 
Science, 348, Issue 6232, pp. 311-314

\bibitem[\protect\citeauthoryear{McKinney}{2005}]{McKinney} 
McKinney J.~C., 2005, ApJ, 630, L5

\bibitem[\protect\citeauthoryear{McKinney \& Blandford}{2009}]{McKinneyBlandford09}
McKinney, J. C. \& Blandford, R. D. 2009, MNRAS, 394L, 126

\bibitem[\protect\citeauthoryear{McKinney et al.}{2012}]{McKinney2012} 
McKinney J.~C., Tchekhovskoy A., Blandford R.~D., 2012, MNRAS, 423, 3083

\bibitem[\protect\citeauthoryear{Meliani \& Keppens}{2010}]{MelianiKeppens10} 
Meliani Z., Keppens R., 2010, A\&A, 520, L3 

\bibitem[\protect\citeauthoryear{Meliani et al.}{2004}]{Melianietal04}
Meliani, Z., Sauty, C., Tsinganos, K., \& Vlahakis, N.  2004, A\&A, 425, 773

\bibitem[\protect\citeauthoryear{Meliani et al.}{2006}]{Melianietal06}
Meliani, Z., Sauty, C., Vlahakis, N., Tsinganos, K. \& Trussoni, E. 2006, A\&A, 447, 797

\bibitem[\protect\citeauthoryear{Meliani et al.} {2010}]{Melianietal10}
Meliani, Z., Sauty, C., Tsinganos, K., Trussoni, E. \& Cayatte, V.  2010, A\&A, 521, 67

\bibitem[\protect\citeauthoryear{Mertens et al.} {2016}]{Mertens16}
Mertens, F., Lobanov, A. P.,  Walker, R. C. and Hardee, P. E. 2016, A\&A, 595, A54

\bibitem[\protect\citeauthoryear{Mignone et al.}{2005}]{2005ApJS..160..199M} Mignone A., Plewa T., Bodo G., 2005, ApJS, 160, 199

\bibitem[\protect\citeauthoryear{Nagai et al.}{2014}]{Nagai14} 
Nagai H., et al., 2014, ApJ, 785, 53

\bibitem[\protect\citeauthoryear{Narayan \& Yi}{1994}]{Nar94} 
Narayan R. \& Yi I., 1994, ApJ, 428, L13

\bibitem[\protect\citeauthoryear{Narayan et al.}{2003}]{Nar03} 
Narayan R., Igumenshchev I.V.,  \& Abramowicz M. A., 2003, PASJ, 55, L69

\bibitem[\protect\citeauthoryear{Nathanail \& Contopoulos}{2014}]{Nathanail14} 
Nathanail A., Contopoulos I., 2014, ApJ, 788, 186 

\bibitem[\protect\citeauthoryear{Parker}{1957}]{Parker57} Parker, E.N., 1957, Physical Review, 107, 924

\bibitem[\protect\citeauthoryear{Parker}{1963}]{Parker63} Parker, E. N. 1963, Interplanetary Dynamical Processes (New York: Interscience Publishers Division, John
Wiley and Sons)

\bibitem[\protect\citeauthoryear{Petropoulou \& Dermer} {2016}]{Petropoulou16}
Petropoulou, M. \& Dermer, C. 2016, ApJL, 825, 11

\bibitem[\protect\citeauthoryear{Prieto et al.}{2016}]{Prieto16} 
Prieto M.~A., Fern{\'a}ndez-Ontiveros J.~A., Markoff S., Espada D., Gonz{\'a}lez-Mart{\'{\i}}n O., 2016, MNRAS, 457, 3801
  
\bibitem[\protect\citeauthoryear{Sauty \& Tsinganos} {1994}]{ST94}
Sauty, C. \& Tsinganos K., 1994, A\&A, 287, 893

\bibitem[\protect\citeauthoryear{Sauty et al.} {1999}]{STT99} 
Sauty, C., Tsinganos, K. \& Trussoni E., 1999, A\&A, 348, 327

\bibitem[\protect\citeauthoryear{Sauty et al.}{2002}]{Sautyetal02} 
Sauty C., Trussoni E., Tsinganos K., 2002, A\&A, 389, 1068 

\bibitem[\protect\citeauthoryear{Sauty et al.}{2004}]{Sautyetal04} 
Sauty C., Trussoni E., Tsinganos K., 2004, A\&A, 421, 797 

\bibitem[\protect\citeauthoryear{Sauty et al.}{2005}]{Sautyetal05} Sauty C., Lima J.~J.~G., Iro N., Tsinganos K., 2005, A\&A, 432, 687 

\bibitem[\protect\citeauthoryear{Sikora} {2011}]{Sikora11} 
Sikora, M. 2011,  in IAU Symp. 275, Jets at all Scales, ed. G. Romero, R.
Sunyaev \& T. Bellon (Cambridge: Cambridge Univ. Press), 59

\bibitem[\protect\citeauthoryear{Sikora et al.}{2016}]{Sikora16} 
Sikora M., Rutkowski M., Begelman M.~C., 2016, \mnras, 457, 1352

\bibitem[\protect\citeauthoryear{Silvestro et al.}{1987}]{Silvestroetal87}
Silvestro, G.,  Ferrari, A., Rosner, R., Trussoni, E. \& Tsinganos, K. 1987, \mnras, 325, 228

\bibitem[\protect\citeauthoryear{Sol et al.}{1989}]{Soletal89} 
Sol, H., Pelletier, G. \& Asseo, E. 1989, \mnras, 237, 411

\bibitem[\protect\citeauthoryear{Taub}{1948}]{1948PhRv...74..328T} Taub A.~H., 1948, PhRv, 74, 328 

\bibitem[\protect\citeauthoryear{Tavecchio} {2016}]{Tavecchio16} 
Tavecchio, F.,  2016, in "Gamma 2016", Heidelberg July 11-15, 2016arXiv160904260T

\bibitem[\protect\citeauthoryear{Tavecchio \& Ghisellini} {2016}]{TavecchioGhis16}
Tavecchio, F.  \& Ghisellini, G. 2016, \mnras, 456, 2374

\bibitem[\protect\citeauthoryear{Tchekhovskoy} {2015}]{Tchek15}
Tchekhovskoy, A. 2015, in Astrophysics and Space Science Library, Vol. 414, 
The Formation and Disruption of Black Hole Jets, ed. I. Contopoulos, D. Gabuzda \& N. Kylafis, 45 

\bibitem[\protect\citeauthoryear{Tchekhovskoy et al.}{2011}]{Tchek2011} 
Tchekhovskoy A., Narayan R., McKinney J.~C., 2011, MNRAS, 418, L79

\bibitem[\protect\citeauthoryear{Tchekhovskoy \& McKinney}{2012}]{Tchek2012} 
Tchekhovskoy A., McKinney J.~C., 2012, MNRAS, 423, L55

\bibitem[\protect\citeauthoryear{Tsinganos} {(1982)}]{Tsing82}
Tsinganos, K. 1982, ApJ, 252, 775

\bibitem[\protect\citeauthoryear{Tsinganos \& Trussoni} {1991}]{TT91}
Tsinganos K., Trussoni, E., 1991, A\&A, 249, 156 

\bibitem[\protect\citeauthoryear{Tsinganos \& Sauty} {1992a}]{TS92a}
Tsinganos K., Sauty, C., 1992a, A\&A, 255, 405 

\bibitem[\protect\citeauthoryear{Urry \& Padovani}{1995}]{UrryPadovani95}
Urry \& Padovani, 1995, PASP, 107, 803

\bibitem[\protect\citeauthoryear{van Eerten et al.}{2011}]{vanEertenetal11} 
van Eerten H.~J., Meliani Z., Wijers R.~A.~M.~J., Keppens R., 2011, MNRAS, 410, 2016 

\bibitem[\protect\citeauthoryear{Vlahakis} {2015}]{Vlah15}
Vlahakis, N., 2015, in The Formation and Disruption of Black Hole Jets, eds 
Contopoulos, I., Gabuzda, D. and Kylafis, N., Astrophysics and Space Science Library, 414, 177

\bibitem[\protect\citeauthoryear{Vlahakis \& K\"onigl} {2003a}]{VlahakisKonigl03a}
Vlahakis, N.\& K\"onigl, A., 2003a, ApJ, 596, 1104

\bibitem[\protect\citeauthoryear{Vlahakis \& K\"onigl} {2003b}]{VlahakisKonigl03b}
Vlahakis, N.\& K\"onigl, A., 2003b, ApJ, 596, 1080

\bibitem[\protect\citeauthoryear{Vlahakis \& K\"onigl} {2004}]{VlahakisKonigl04}
Vlahakis, N.\& K\"onigl, A., 2004, ApJ, 605, 656

\bibitem[\protect\citeauthoryear{Vlahakis \& Tsinganos} {1998}]{VlahakisTsinganos98}
Vlahakis, N.\& Tsinganos, A., 1998, \mnras, 298, 777

\bibitem[\protect\citeauthoryear{Vovk \& Babi{\'c}}{2015}]{Vovk15} 
Vovk I., Babi{\'c} A., 2015, A\&A, 578, A92

\bibitem[\protect\citeauthoryear{Zamaninasab et al.}{2014}]{Zama2014} 
Zamaninasab M., Clausen-Brown E., Savolainen T., Tchekhovskoy A., 2014, Natur, 510, 126 

\bibitem[\protect\citeauthoryear{Zdziarski et al.}{2015}]{Zdziarski2015} 
Zdziarski A.~A., Sikora M., Pjanka P., Tchekhovskoy A., 2015, MNRAS, 451, 927

\end{thebibliography}
\end{document}